%% file: main.tex
\documentclass[10pt, sigconf,authorversion,nonacm]{acmart}

\setcopyright{none}

\thanks{\textcolor{red}{This paper has been accepted by the ACM Symposium on Access Control Models and Technologies (SACMAT) 2022. The definite version of this work will be published by ACM as part of the SACMAT conference proceedings.}}

\usepackage[utf8]{inputenc}
\usepackage{enumitem}
\usepackage{fixltx2e}
\usepackage{graphicx}
\usepackage{epsfig}
\usepackage{longtable}
\usepackage{float}
\usepackage{dblfloatfix}
\usepackage{wrapfig}
\usepackage{soul}
\usepackage{color}
\usepackage{textcomp}
\usepackage{marvosym}
\usepackage{wasysym}
\usepackage{latexsym}
\usepackage{url}
\usepackage{subcaption}
\usepackage{multirow}

\usepackage{listings}
\usepackage{enumitem}
\usepackage{verbatim}
\usepackage{seqsplit}
\usepackage{multirow}
\usepackage{array}
\setlist{nolistsep}  
\usepackage[font=scriptsize,margin=.5cm]{caption}
\usepackage{color,soul}
\usepackage{xspace}
\usepackage{calc}
\usepackage{algorithm}
\usepackage{algorithmicx}
\usepackage{algcompatible}
\usepackage{enumitem}
\usepackage[compact]{titlesec}

\titlespacing{\subsubsection}{0pt}{*0}{*0}

\newcommand{\spyros}{\color{black}}

\newcommand{\reza}{\color{black}}

\newcommand{\sol}{{\em Harpocrates}\xspace}

\newcommand{\pull}{{\em Pull}\xspace}
\newcommand{\push}{{\em Push}\xspace}
\newcommand{\hybrid}{{\em Hybrid}\xspace}

\newcommand{\ie}{{\em i.e.,}\ }
\newcommand{\eg}{{\em e.g.,}\ }

\newcommand{\collabPeers}{collaborating peers\xspace}
\newcommand{\collabProxies}{collaborating proxies\xspace}
\newcommand{\censorNodes}{censoring nodes\xspace}

\newcommand{\selProxies}{selected proxy\xspace}

\newcommand{\collabPeer}{collaborating peer\xspace}
\newcommand{\collabProxy}{collaborating proxy\xspace}

\newcommand{\selProxy}{selected proxy\xspace}

\newtheorem{definition}{\bf Definition}   [section]
\newcommand{\kENDdef}   {$\Box$}		
\newcommand{\bZ} {{\mathbb{Z}}}
\newcommand{\bG} {{\mathbb{G}}}

\floatname{algorithm}{Protocol}

\usepackage{etoolbox}
\usepackage{xstring}
\DeclareListParser{\doslashlist}{/}
\newcounter{ndnNameComponentCounter}%
\newcommand{\name}[1]{{%
  \setcounter{ndnNameComponentCounter}{0}%
  \renewcommand{\do}[1]{{%
    \ifnumgreater{\value{ndnNameComponentCounter}}{0}{\allowbreak/}{}%
    \ifnumodd{\value{ndnNameComponentCounter}}{}{}%
    ##1}%
    \stepcounter{ndnNameComponentCounter}}%
``{\fontfamily{cmtt}\small\selectfont\IfBeginWith{#1}{/}{/}{}\doslashlist{#1}}''%
}}

\usepackage[compact]{titlesec}
\titlespacing{\section}{0pt}{2pt}{2pt}
\titlespacing{\subsection}{0pt}{2pt}{2pt}

\makeatletter
\def\@copyrightpermission{Test}
\makeatother

\author{Md Washik Al Azad}
\affiliation{University of Nebraska at Omaha}
\email{malazad@unomaha.edu}
\author{Reza Tourani}
\affiliation{Saint Louis University}
\email{reza.tourani@slu.edu}
\author{Abderrahmen Mtibaa}
\affiliation{University of Missouri--Saint Louis}
\email{amtibaa@umsl.edu}
\author{Spyridon Mastorakis}
\affiliation{University of Nebraska at Omaha}
\email{smastorakis@unomaha.edu}

\begin{document}

\title{\sol: Anonymous Data Publication in Named Data Networking}


\begin{abstract}
Named-Data Networking (NDN), a prominent realization of the Information-Centric Networking (ICN) vision, offers a request-response communication model where data is identified based on application-defined names at the network layer. This amplifies the ability of censoring authorities to restrict user access to certain data/websites/applications and monitor user requests. The majority of existing NDN-based frameworks have focused on enabling users in a censoring network to access data available outside of this network, without considering how data producers in a censoring network can make their data available to users outside of this network. This problem becomes especially challenging, since the NDN communication paths are symmetric, while producers are mandated to sign the data they generate and identify their certificates. In this paper, we propose \sol, an NDN-based framework for anonymous data publication under censorship conditions. \sol enables producers in censoring networks to produce and make their data available to users outside of these networks while remaining anonymous to censoring authorities. Our evaluation demonstrates that \sol achieves anonymous data publication under different settings, being able to identify and adapt to censoring actions.
\end{abstract}

%
%




\maketitle

\input{Sections/intro.tex}

\input{Sections/background.tex}

\input{Sections/model.tex}

\input{Sections/overview.tex}

\input{Sections/security.tex}

\input{Sections/design.tex}

\input{Sections/eval.tex}

\input{Sections/discussion.tex}

\input{Sections/conclusion.tex}

\section*{Acknowledgements}

This work is partially supported by National Science Foundation awards CNS-2104700, CNS-2016714, and CBET-2124918, the National Institutes of Health (NIGMS/P20GM109090), the Nebraska University Collaboration Initiative, the Nebraska Tobacco Settlement Biomedical Research Development Funds, and Intel Labs through a gift.

\bibliographystyle{acm}
\bibliography{Sections/refs}

\end{document}

%% file: Sections/intro.tex
\section{Introduction}
\label{sec:intro}


Preserving the anonymity of users that generate and share data (\eg pictures, videos, messages) with others is crucial especially in scenarios where the safety and freedom of users may be in danger (\eg authoritarian regimes)~\cite{shklovski2011online}. In such scenarios, authorities, such as governments, Internet Service Providers (ISPs), and other organizations, may restrict access to certain websites and block the operation of applications that allow users to publish their data on the Internet~\cite{leberknight2010taxonomy}. The ultimate goal of these authorities is to limit access to data that they do not consider favorable and avoid having non-favorable data (\eg videos of protests, pictures of illegal practices) be published by their users on the Internet. For example, during protests in a certain country, protesters may take pictures or videos that show law enforcement personnel attempting to violently suppress these protests. The government may restrict protesters from uploading this data to popular hosting (\eg YouTube and Vimeo), news (\eg CNN and BBC), and social media (\eg Facebook and Instagram) websites. Even in cases that users find ways to upload their data on the Internet (\eg on websites not blocked by the government), the government in cooperation with local ISPs may be able to identify the citizen(s) that uploaded the data and imprison them. At the same time, hosting, news, and social media websites have vested interest in verifying the authenticity of the uploaded data before making it available to their users without compromising the anonymity of the data producer.

This scenario highlights the following fundamental questions when it comes to publishing data under censorship: 
{\it (i) how can citizens/users that produce data within oppressive countries and organizations (censoring networks) publish this data on the public Internet (non-censoring networks)?
and (ii) how can the produced data be published and authenticated on the public Internet while its producers remain anonymous to the oppressive countries and organizations, which could threaten the producers' safety and well-being?} Solutions to tackle these issues have been proposed in the context of the IP-based network architecture~\cite{Tor, winter2012great, winter2016identifying, durumeric2013zmap,zolfaghari2016practical,karlin2011decoy}.



Over the last decade, the direction of Information-Centric Networking (ICN)~\cite{ahlgren2012survey} and its prominent realization, Named-Data Networking (NDN)~\cite{zhang2014named}, have attracted attention by the research community. NDN features a request-response communication model, where requests that identify the data by application-defined names are forwarded towards data producers. {\spyros NDN possesses the privacy friendly features of not containing specific source and destination addresses in its packets. However, the use of semantically rich names at the network layer amplifies the ability of censoring authorities to restrict access to non-favorable data and monitor what data their users request.}


Several solutions have been proposed to alleviate this issue~\cite{TouMicMis16, AriKop12, DibGasTsu11, BerMarasg19, MozHouVen19}, focusing on how users in a censoring network can access data available in an non-censoring network. However, these solutions did not consider how producers residing in a censoring network can make their data available to users outside of this network. This problem in NDN is especially challenging due to: (i) the retrieval process, initiated by requests that carry the names of the data to be retrieved, empowers censoring authorities to drop requests for data produced within the censoring network at the border of this network; (ii) the symmetry of the communication model (\ie each response follows the same network path back to a requester as the corresponding request) enables censoring authorities to analyze requests and block the corresponding data on the way back to the requester; and (iii) as a by-product of (i), (ii), and the NDN semantically meaningful naming, producers in censoring networks cannot advertise the data they produce, since this would enable censoring authorities to directly link the generated data to them and cannot be reachable from outside of censoring networks, since censoring authorities can easily drop incoming requests with non-favorable or unknown data names.


To tackle these challenges, we propose \sol\footnote{\sol was the god of secrets and confidentiality according to the ancient Greek mythology.}, an NDN-based framework for anonymous data publication, which enables producers in censoring networks to produce, upload, and make their data available to users outside of these networks while remaining anonymous to censoring authorities. \sol makes the following contributions: 

\begin{itemize}[leftmargin=*, topsep=0pt] 

\item It takes advantage of communication channels and applications that operate legally within the censoring network as well as a decoy routing approach~\cite{karlin2011decoy} to publish data in a peer-to-peer fashion, maximizing the collateral damage for censoring authorities; 

\item It features mechanisms for producers to identify censoring activities and adapt their data publication process to such activities. This ensures that the data will be successfully uploaded to a network of trusted proxies in order to become available to users outside of the censoring network;

\item It features a secure delegation mechanism between producers and proxies, preventing censoring authorities {\spyros from being able to link the generated data back to producers.} 
As a result, proxies can make the data available outside of the censoring network on behalf of producers while preserving the producers' anonymity and, at the same time, enabling users to verify data authenticity.

\end{itemize}

To the best of our knowledge, \sol is among the first attempts in NDN/ICN environments to tackle the problem of making data generated within a censoring network available outside of this network without compromising the producer's anonymity. 

%% file: Sections/background.tex
\section{Background and Prior Work}
\label{sec:background}

In this section, we give a brief background of the NDN architecture and discuss related work in both IP and NDN/ICN environments.

\subsection {Named-Data Networking}

NDN~\cite{zhang2014named} features a receiver-driven model that leverages application-defined semantically meaningful naming for communication purposes. In NDN, \emph{consumer applications} send requests for data, called Interest packets. Interests are forwarded based on their names towards \emph{data producer applications}, which send Data packets that contain the requested data back to consumers. 

For the realization of the NDN communication model, NDN routers maintain three data structures: (i) a Forwarding Information Base (FIB), which consists of name prefixes along with a number of outgoing interfaces for each prefix and is used for Interest forwarding; (ii) a Pending Interest Table (PIT), which stores Interests that have been recently forwarded but have not retrieved data yet; and (iii) a Content Store (CS), where retrieved Data packets are cached to satisfy future requests for the same data. 

NDN is based on three fundamental principles: (i) \emph{identifying network-layer packets through application-defined, semantically meaningful names}--NDN carries network-layer packets that contain application-defined names; (ii) \emph{securing data directly at the network layer}--each network-layer Data packet carries the signature of its producer, which cryptographically binds the actual data to the packet's name and secures the data at rest and in transit across the network, along with signature related information that specifies the producer's certificate or public key~\cite{ndn2015ndn}
; and (iii) \emph{a stateful forwarding plane:} forwarded Interests leave state at each router, while Data packets follow the reverse path of the corresponding Interests, consuming the state at each router.

\subsection {Prior Work on Censorship Circumvention and Anonymity}
\label{sec:background}
\label{subsec:related-work}




\subsubsection {IP-based Censorship Circumvention and Anonymity}

Tor~\cite{Tor} is the most popular anonymity network, which uses an overlay of relays to provide identity anonymity and unlinkability. Extensive research has been conducted on various facets of Tor~\cite{winter2012great, winter2016identifying, durumeric2013zmap}. However, using layers of encryption and decryption to secure the data imposes considerable overhead and impacts communication latency. 
Tor's high latency and its vulnerability to active probing~\cite{EnsWinMue15} motivated the design of an alternative approach, decoy routing~\cite{karlin2011decoy}. Decoy routing is an in-network censorship circumvention platform, where a set of decoy routers participate in relaying the traffic outside of a censoring network. Several flavors of decoy routing have been proposed to enhance the seminal design through decoy placement optimizations~\cite{schuchard2012routing} 
routing optimizations based on game theory~\cite{nasr2019enemy}, mimicking access patterns to non-censored websites~\cite{bocovich2016slitheen}, and routing asymmetries~\cite{nasr2017waterfall}. Another censorship evading direction includes mimicking the traffic profiles of non-censored, innocuous applications~\cite{wustrow2011telex, fifield2015blocking}. The community has also investigated frameworks that utilize public Content Delivery Networks (CDNs) to access censored data under the assumption that blocking data hosted on these CDNs will cause collateral damage, since innocuous data publishers will be disrupted~\cite{zolfaghari2016practical}. 






\subsubsection {NDN/ICN-based Censorship Circumvention and Anonymity}

The state-of-the-art in NDN/ICN censorship circumvention and anonymous communication is categorized into proxy-independent and proxy-based techniques~\cite{TouMicMis16}.
In this realm, the use of steganography, where data and a cover file need to be combined before publication, was among the first proposals~\cite{AriKop12}.
Users obtain the necessary data decoding information through a secure back channel. This scheme imposes considerable communication overhead, which impacts its scalability.
Techniques, such as homomorphic encryption, have been also proposed in a publish-subscribe design 
to provide privacy for user requests~\cite{FotTroMar14}. 
%

%
Tor has inspired proxy-based solutions~\cite{DibGasTsu11, kita2020producer}, where layers of encryption between users and a network of proxies are used for anonymity.
CoNaP~\cite{LesYaqKha19} takes a similar approach, where a user encrypts and signs the names of Interests for authenticity. However, this signature reveals the user identity and compromises its anonymity~\cite{ramani2019ndn}.
To reduce the cost of a symmetric key cryptosystem, which needs to be carried on a per-packet basis, lightweight coding techniques, including random linear network coding~\cite{TaoFeiYe15} and Huffman coding~\cite{TouMisKli15}, have been proposed.
PrivICN~\cite{BerMarasg19} is another proxy-based scheme that enables cache utilization. By employing proxy re-encryption, PrivICN enables cached data in the censoring network to be used by multiple users. However, cache hits in the censoring network introduce information leakage and undermine user anonymity.
A decoy routing approach was also proposed for traffic redirection~\cite{MozHouVen19}, where a user informs a decoy router through a covert channel to redirect its requests to the covert rather than the decoy destination.
Finally, an Attribute-Based Signature scheme for NDN was proposed~\cite{ramani2019ndn}. However, this scheme focuses on anonymizing a producer's signatures, without considering any other aspects of the anonymous data publication process.
 
\noindent \textbf{How does \sol differ from prior work?} While the majority of existing NDN/ICN approaches have focused on enabling consumers within a censoring network to reach producers in non-censored networks to download data, very few designs have considered the problem of anonymous data publication in NDN/ICN. Such designs primarily focus on signature anonymization~\cite{ramani2019ndn} or rely on onion routing, which requires multiple, costly layers of encryption~\cite{kita2020producer}. Our work enables producers in a censoring network to publish (upload) their data to consumers outside of this network in an anonymous manner without the need for multiple layers of encryption.

%

%% file: Sections/model.tex
\section{Model and Assumptions}
\label{sec:model}

In this section, we present our system and network model, our design assumptions, our threat model, and the goals of the \sol design. Table~\ref{notations} includes the notations we use in the rest of this paper.
\begin{table}[t]
\centering
\vspace{-0.1cm}
\scriptsize
\caption{Summary of notations.}
\vspace{-0.3cm} 
\label{notations}
\begin{tabular}{|c|p{2in}|}
 \hline
 	Notation & Description \\
 \hline
	$P$, $Q$ & Big prime numbers such that $P=2Q+1$\\
	$\bZ_{Q}^{*}, \bZ_{P}^{*}$ & Multiplicative groups of integers of
 	 order $Q$ and $P$ respectively \\
	$\bG_Q, \bG_P$ & Cyclic groups of order $Q$ and $P$ respectively \\
 	$\bG_S$ & Schnorr group (large prime-order subgroup of $\bZ_{P}^{*}$) \\
	$g$ & Generator of a sub-group of $\bG_P$ of order $Q$ \\
	$ZQrand()$ & Random number generator in $\bZ^{*}_Q$ \\
	$(PK_X,PR_X)$ & X's public and private signing key pair \\
	$S_{XY}$ & Symmetric key shared between $X$ and $Y$ \\ 
	$H():\{0,1\}^{*}\rightarrow \bZ^{*}_Q$ & Cryptographic hash function with digest $\in \bZ^{*}_Q$\\
	$W$ & Warrant for proxy signature delegation \\
	$M$ & Message to be signed \\
	$||$ & Concatenation operator \\
	$\equiv$ & Congruence operator \\
 \hline
\end{tabular}
\end{table}

\subsection{System and Network Model}
\label{subsec:systemmodel}

\begin{figure}[t]
\centering
\vspace{-0.2cm}
\includegraphics[width=0.78\columnwidth]{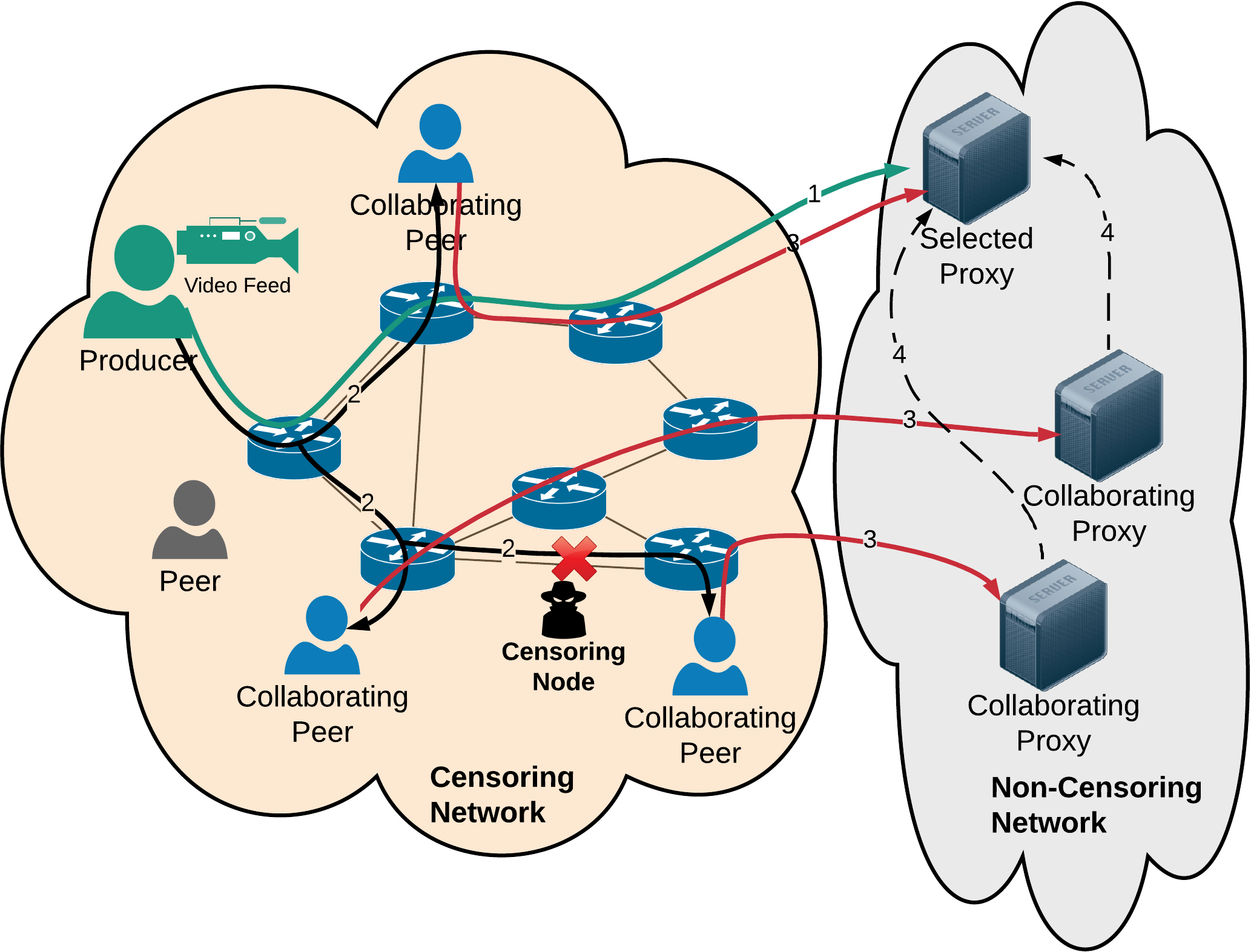}
\vspace{-0.4cm}
\caption{System model of \sol: (1) the producer establishes a covert channel with a \selProxy; (2) the producer shares pieces of data with \collabPeers in a peer-to-peer fashion, while \censorNodes may intercept these pieces; (3) \collabPeers push the data towards \collabProxies outside of the censoring network in manners that prevent traffic analysis attacks; and (4) \collabProxies share the data with the \selProxy that makes it available to consumers outside of the censoring network. }
\label{fig:system}
\end{figure}

We consider a censoring network and a set of proxies that make data available to consumers outside of this network (Fig.~\ref{fig:system}). Our system model consists of the following actors: 
\begin{itemize}[leftmargin=*]

\item \textbf{Producer:} An entity in the censoring network that wishes to anonymously publish data (potentially consisting of several network-layer Data packets) outside of this network. 
\item \textbf{Consumer:} An entity outside of the censoring network interested in the data generated by the producer. 

\item \textbf{Peers:} Entities (in the censoring network) subscribed to a peer-to-peer application that operates ``legally'' in the censoring network. The producer is a peer running this application. 

\item \textbf{Collaborating peers:} Peers selected by the producer 
to help make the data generated by the producer available outside of the censoring network.

\item \textbf{Censoring nodes:} Entities deployed by ISPs, governments, or other stakeholders in the censoring network to detect and block attempts to publish data outside of this network. 

\item \textbf{Selected proxy:} A trusted entity outside of the censoring network that collects, reconciles, and publishes the data on behalf of the anonymous producer, so that consumers outside of the censoring network can access this data. 

\item \textbf{Collaborating proxies:} Trusted entities outside of the censoring network that receive censored data from the \collabPeers and send this data to the \selProxies.

\end{itemize}

We illustrate our system through a running example in Fig.~\ref{fig:system}. The producer selects a set of \collabPeers (subset of the overall peers) and shares with them pieces of the generated data. However, these pieces can be intercepted and blocked by \censorNodes on their way to the \collabPeers. A \collabPeer receiving a data piece will send it towards a \collabProxy. 
Each \collabProxy will eventually forward the received data to the \selProxy. 

\subsection{Assumptions}
\label{subsec:assumptions}
We assume that producers in the censoring network do not advertise their data to protect their anonymity. We also assume that producers are not reachable from outside of the censoring network, thus they cannot directly upload their data to consumers outside of this network. This is a fair assumption considering the symmetric, name-based nature of NDN communication. This makes it trivial for censors to block requests or responses for data produced in their network and for entities within their network to ensure that censored data does not become available to the outside world.


%
We consider rational attackers with bounded capabilities, that is attackers who do not orchestrate large-scale brute force attacks or block all the communication in the censoring network. This is a fair assumption since pervasive blocking causes collateral damage~\cite{zolfaghari2016practical}. 
We assume that neither \collabProxies nor \collabPeers (selected by the producer) are malicious. This is a fair assumption since the majority of censorship circumvention tools leverage trust and reputation-based mechanisms to select entities playing key roles. For instance, in Tor, only trustworthy relay nodes can be selected as {\it entry guards} due to their importance in protecting user anonymity~\cite{ElaBayAls12}.
{\reza We assume the existence of an anonymous public-key certificate approach~\cite{huang2010anonymous}, which preserves the privacy of the producers' information in their certificates. We discuss directions to further augment producer anonymity in Section~\ref{sec:discussion}.}
Finally, we assume that symmetric and asymmetric cryptographic operations are secure.

\vspace{-0.1cm}
\subsection {Threat Model}
\label{subsec:threatmodel}
In NDN, the use of names at the network layer can simplify data filtering, censorship, and violate the consumer and producer privacy. In this paper, we consider that a censoring authority can deploy active attackers and passive eavesdroppers across the censoring network to interrupt ongoing data publications from this network to the outside world or compromise producer anonymity. An active attacker can capture and modify transmitted packets, while a passive eavesdropper can analyze the captured packets. Deployed attackers may masquerade as different entities such as peers.

The primary objective of the censoring authority is to prevent producers in the censoring network from publishing data.
Thus, the censoring authority may:
{\it (i)} block the ISP's ingress Interests destined to producers; 
{\it (ii)} act as a man-in-the-middle to collect the requested data from the producers, compare it against a blacklist, and either drop the packets or relay them to the requester;
{\it (iii)} deploy \censorNodes to interrupt the ongoing communication across peers by dropping the Interest and/or Data packets; and
{\it (iv)} masquerade as a peer 
to interrupt the communication and compromise the producer's anonymity.
We note that objectives {\it (iii)} and {\it (iv)} are different in the sense that in the former one, the attacker is an ISP node in the censoring network (\eg a router), while in the latter one, the attacker is one of the peers.
%
While the focus of this work is enabling anonymous data publication rather than coping with traffic analysis attacks, we will briefly discuss potential traffic analysis countermeasures in Section~\ref{sec:discussion} to thwart this category of attacks.

\subsection {\sol Design Goals}
\label{subsec:designgoals}
%
%
%
\sol offers data producers--whether in a censoring network or not--to successfully publish their data (evade censorship) while preserving their privacy and data integrity.
\sol has the following goals:

\begin{itemize} [wide, labelwidth=!, labelindent=0pt, nosep]

\item {\bf Anonymity and plausible deniability:} \sol should preserve the producer's anonymity in the presence of different attackers. 
The attackers may interrupt the data publication, but should neither be able to reveal the producer's identity {\spyros nor link the published data to the producer.} 

\item {\bf Integrity guarantees:} \sol should guarantee the integrity of the published data without revealing the producer's identity. This is important as the producer delegates the publication of its data to a third party (\selProxies).

\item {\bf Reasonable overhead:} \sol should incur reasonable communication and computation overhead on the involved actors.
The cost of \sol for the \collabPeers should be viable, while the producer should be able to publish its data with reasonably low latency.
\end{itemize}

%% file: Sections/overview.tex
\section{Design Overview}
\label{sec:overview}
%

\begin{figure}[!t]
\centering
\includegraphics[width=0.9\columnwidth]{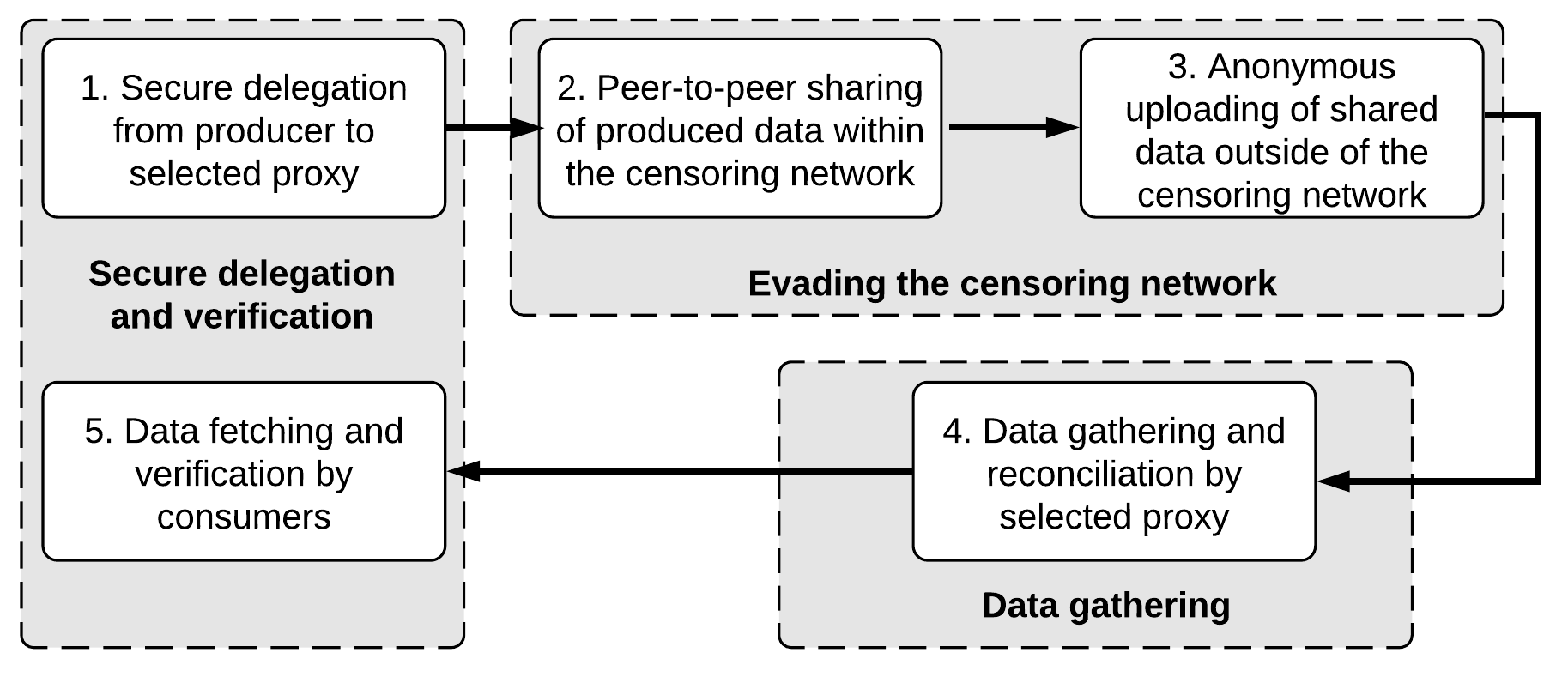}
\vspace{-0.4cm}
\caption{Overview of the \sol design.}
\label{fig:overview}
\end{figure}

In this section, we present an overview of \sol (Fig.~\ref{fig:overview}). In \sol, the producer will first reach and securely delegate the data publication privilege to the \selProxy that will help preserve the producer anonymity. After the secure delegation phase, the producer will start the data uploading phase through a peer-to-peer mechanism to: {\spyros (i) prevent the censoring authority from detecting abnormal amounts of data from a single peer being sent outside of the censoring network; and (ii) ensure that the data production cannot be linked back to the producer.} 

%

These phases are facilitated through the use of decoy routing techniques~\cite{KarEllJac11}, which we briefly discuss in the rest of this section and provide details in Sections~\ref{sec:security} and~\ref{sec:design}. Different than in IP, decoy routing in \sol is realized through decoy name prefixes (\ie prefixes that allow Interests from within the censoring network to be forwarded outside of this network). Benign information encoded in names help proxies identify Interests with decoy prefixes. Combined with the fact that NDN routers have direct access to the names of Interests, decoy routing in \sol can be easily deployed outside of the censoring network, without requiring routers to search for signalling information at higher layers of the protocol stack (\eg transport or application layer)~\cite{KarEllJac11}. This makes the deployment of decoy routers flexible and simple, while decoy router assignments can change over time (\eg coordinated through routing protocols in non-censoring networks and selected based on placement strategies that maximize collateral damage~\cite{nasr2019enemy}) to overcome known attacks against decoy routing (\eg routing around decoys~\cite{schuchard2012routing}).


\subsection{Secure Delegation Overview}
\label{sigdel}
%
In this phase, the producer employs decoy routing to select and reach one of the \collabProxies, who will act as the \selProxy, outside of the censoring network. Subsequently, the producer securely provides delegation metadata and instructions to the \selProxy for data publication. 
The \selProxy, on receiving the metadata, accepts the data publication by returning a signed {\it commitment} to prove its involvement in data publication and enable the initiation of the proxy signature process (Definition~\ref{defn:proxy}). We use a warrant-based proxy signature~\cite{AboYou12} based on the difficulty of the discrete logarithm problem.
\begin{definition}
\label{defn:proxy}
{\bf [Proxy Signature]} 
Proxy signature is a cooperative digital signing scheme~\cite{MamUsuOka96}, in which an original signer (data producer in our case) delegates its right of digitally signing a message to a proxy. Such a delegation allows verifiers (consumers in our case) with knowledge of the signer's and proxy's public keys to validate signed messages. 
Proxy signature schemes are categorized into full delegation, partial delegation, and delegation by warrant. A warrant includes metadata, such as delegation scope information to authorize the proxy to sign on behalf of the original signer.
\hfill\kENDdef      
\end{definition}

The producer then generates the required credentials for proxy signature and securely sends them to the \selProxy for data signing and publication. In \sol, we use Schnorr group (Definition~\ref{defn:schnorr}) and Schnorr signature~\cite{Sch91}.
\begin{definition}
\label{defn:schnorr}
{\bf [Schnorr Group]} 
Given two large primes $Q$ and $P$, where $P = rQ + 1$, $r \in \bZ^{*}_Q$ and $\bZ^{*}_Q$ is the multiplicative 
group of integers $mod$ $Q$, choose $1 < h < P$, such that $ h^r \not\equiv 1$ $mod$ $P$, then $g = h^r$ generates a Schnorr group ($\bG_S$). $\bG_S$ is a subgroup of $\bZ^{*}_P$, the multiplicative group of integers $mod$ $P$ of order $Q$~\cite{Sch91}. 
\hfill\kENDdef      
\end{definition}

\subsection{Anonymous Data Uploading Overview}
\label{subsec:uploadover}

Overall, the \sol communication design consists of two main steps: (i) {\em evading the censoring network} by sending all producer's Data packets to \collabProxies outside of censoring network without compromising producer's anonymity; and (ii) {\em gathering of all Data packets} by the \selProxy, reconstructing the original producer's data, and making this data available to consumers on the Internet. 

To maximize collateral damage 
for the censoring authority, \sol features a peer-to-peer mechanism, where the producer makes its data available through decoy routing towards proxies outside of the censoring network. Subsequently, Internet users can fetch the uploaded data from the proxies. The peer-to-peer mechanism leverages existing and allowed channels of communication (\eg gaming or local social media applications) in the censoring network. These allowed applications and communication channels are used to ``hide'' data transfers towards the \collabPeers, spreading the data uploading traffic across these peers in the censoring network. We note that having producers use decoy routing to directly reach collaborating proxies would result in significant volumes of traffic initiated by producers and traffic anomalies that can be detected by censors.


In NDN, communication among multiple parties (\eg for a multiplayer gaming application) is realized through a distributed synchronization protocol~\cite{li2018brief}. This protocol creates a multicast name prefix for communication among all the peers in a group (\eg users that play an online game as a team), ensuring that a request sent from one peer in this group will be received by all other peers. Through this synchronization process and the established multicast channel, \collabPeers can share information. However, this process can be infiltrated by the censoring network through the deployment of \censorNodes as routers and/or peers in the synchronization group to intercept traffic\footnote{The routers deployed as \censorNodes not only can intercept but also drop the exchanged traffic. However, this would cause significant collateral damage given that peers utilize allowed communication channels and applications as we further discuss in Section~\ref{sec:discussion}.}. To cope with that, \sol offers a data encryption mechanism during the communication among \collabPeers, ensuring that only selected \collabPeers will be able to decrypt the exchanged information.

%% file: Sections/security.tex
\section{Secure Delegation Design}
\label{sec:security}
\begin{figure}[t]
\centering
\includegraphics[width=1\columnwidth]{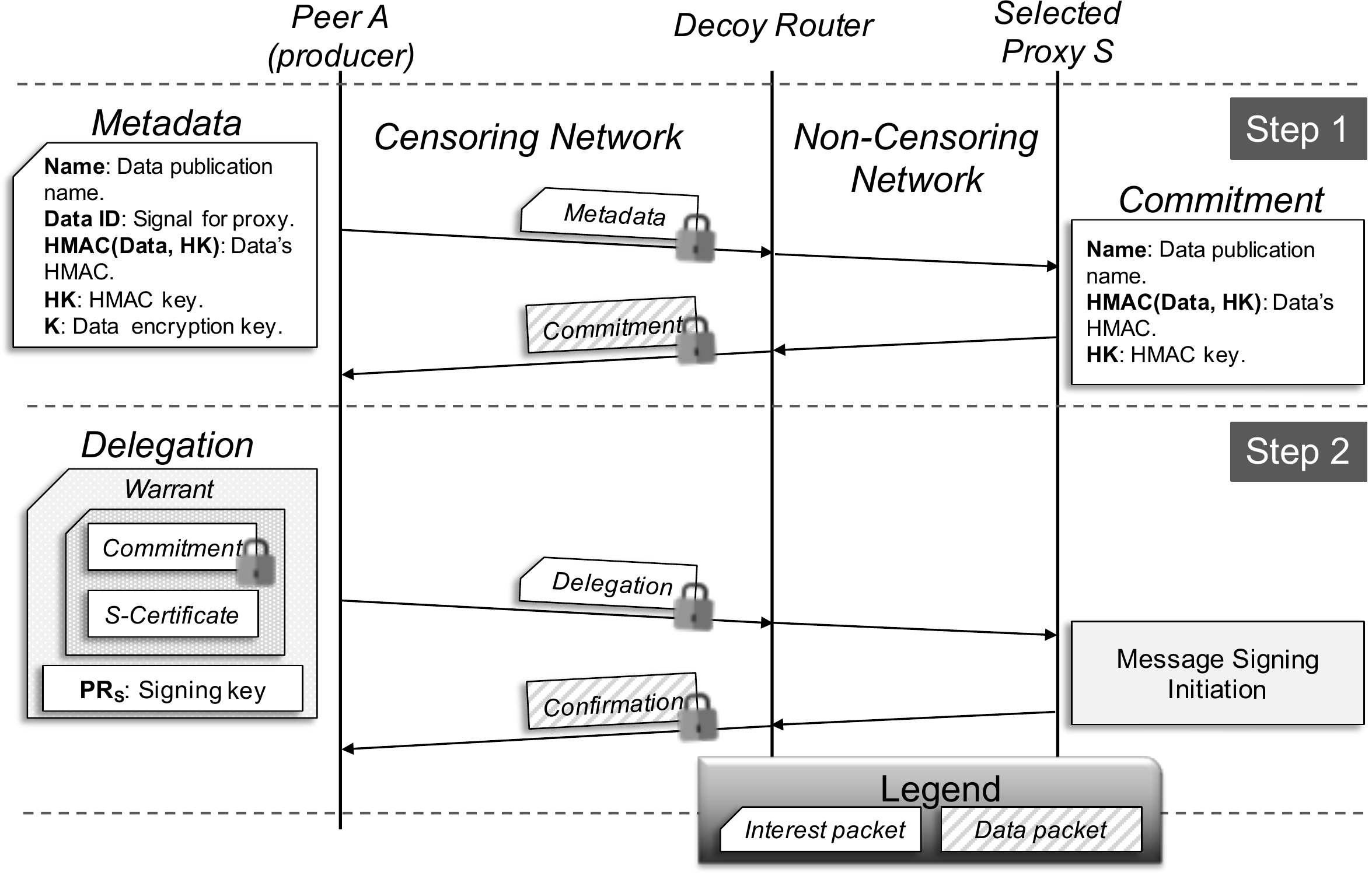}
\vspace{-0.8cm}
\caption{The secure delegation phase of \sol has two steps. First, peer A (producer) provides delegation metadata to the \selProxies (S) for data publication and obtains the \selProxies's commitment. Subsequently, peer A provides the credentials for proxy signature to the \selProxies.}
\label{fig:phase1}
\end{figure}

In this section, we present the secure delegation phase (Fig.~\ref{fig:phase1}), including the {\it proxy commitment}, {\it signature generation}, {\it delegated message signing}, and {\it signature verification} protocols. 
The goal is for the producer to get the selected proxy to commit publishing the data on behalf of the producer, while guaranteeing data integrity, confidentiality, and anonymity.
{\reza As mentioned in Section~\ref{subsec:assumptions}, we assume that producers have anonymous public key certificates~\cite{huang2010anonymous} to avoid revealing information about themselves through their certificates. In Section~\ref{sec:discussion}, we discuss approaches to augment the anonymity level that generic public key certificates provide.}
%
%

\subsection{Proxy Commitment}
As shown in Step~1 of Fig.~\ref{fig:phase1}, the producer generates delegation metadata, including the {\it Data Name} under which the selected proxy should publish the data, the {\it Data ID} to signal the \selProxy of the related Data packets, the data {\it HMAC} (keyed-hash message authentication code) and its key ({\it HK}), and the data encryption key ({\it K}). The Data ID is a random string to be used as part of the data name, informing the \selProxy of the Data packets that belong to this particular data collection. The producer then securely (signed and encrypted) transmits the metadata to the \selProxy. We employ decoy routing to make the proxies reachable to the peers inside the censoring network. As stated in Section~\ref{subsec:threatmodel}, the censoring authority blocks requests from outside of the censoring network to prevent leaking internal data. Thus, the producer and peers can only communicate with proxies by attaching information (\eg delegation metadata) to Interest packets and send them using decoy name prefixes. 

Upon receiving the delegation metadata, the \selProxies (``proxy'' in the rest of this section) uses the metadata to create the {\it commitment}, including the data name, its HMAC, and the HMAC's key. It then signs the commitment using its primary key pair, encrypts it using the shared session key, and returns it to the producer. The rationale behind enforcing the proxy to generate the commitment is to prevent a malicious proxy from altering the producer's data before publication.
The use of warrant-based proxy signatures~\cite{AboYou12} requires the producer to generate a warrant and a signing key pair for the \selProxy--from its own asymmetric key. The generated signing key (delegated key pair) is different from the \selProxy's primary key pair and should be used for proxy signature. 

\subsection{Producer Signature Delegation}
As shown in Step~2 of Fig.~\ref{fig:phase1}, peer $A$ generates a warrant composed of the \selProxy's signed commitment, the proxy's certificate (corresponds to its primary key pair for commitment verification), and the producer's public key (Protocol~\ref{generate}). This information authorizes the \selProxy to sign on behalf of the producer and restricts the \selProxy from abusing the delegated authority (\eg altering the data or its name included in the commitment).
Having the warrant generated, the producer needs to derive the proxy's delegated key pair through the proxy signature scheme. 

\begin{algorithm}[!t]
\caption{Proxy Signature Generation by Producer $A$}
\label{generate}
\scriptsize
\begin{algorithmic}[1]
\REQUIRE{$\bG_S$, $W$, $H()$, and $ZQrand()$.}
\ENSURE{$PR_S$ (Proxy $S$ private key).}
\STATE Choose signing private key $PR_A \in \bZ^{*}_Q$. 
\STATE Calculate corresponding public key $PK_A = g^{PR_A} \in \bZ^{*}_P$.
\STATE Select $i = ZQrand()$.
\STATE Calculate $t = g^i \in \bZ^{*}_P$.
\STATE Generate $W_h = H(W||t)$.
\STATE Calculate $PR_S = (W_h \times PR_A + i) \in \bZ^{*}_Q$.
\STATE Store $<PR_S, W, t>$
\end{algorithmic}
\end{algorithm}
%

Protocol~\ref{generate} takes an agreed upon Schnorr group ($\bG_S$), the hashing function ($H()$), and the warrant ($W$) as inputs and returns the private signing key ($PR_S$) of the \selProxy (proxy $S$); the delegated private key. Producer $A$ initiates this process by choosing a Schnorr private signing key ($PR_A$) and generating the corresponding public verification key ($PK_A$) (Lines~1-2).
To derive the \selProxy's signing key, $A$ selects a random integer $i$ in the multiplicative groups of integers of order $Q$ and calculates $t$ (Lines~3-4). It then generates the warrant's digest ($W_h$) using $W$ and $t$ (Line~5). In Line~6, $A$ uses the warrant's digest ($W_h$), its private key ($PR_A$), and $i$ to calculate the \selProxy's private signing key ($PR_S$). The equation in Line~6 shows the involvement of $A$'s private key in generating the \selProxy's private signing key. Finally, $A$ securely sends the generated private key ($PR_S$), the warrant ($W$), and $t$ to $S$. The completion of Protocol~\ref{generate} concludes the interactions between $A$ and $S$. 

\subsection{Proxy Data Signing}

Upon collecting all Data packets, proxy $S$ executes Protocol~\ref{sign} to sign the packets on behalf of producer $A$ using the delegated private signing key ($PR_S$). Protocol~\ref{sign} accepts the agreed upon Schnorr group ($\bG_S$), the hash function ($H()$), a Data packet (message $M$), and the three-tuple $S$ received from $A$ ($<PR_S, W, t>$) and returns a signed packet.

Initially, $S$ uses $\bG_S$ and its private signing key ($PR_S$) to generate the corresponding public verification key ($PK_S$), which will be used by consumers to verify the delegated proxy signature on the Data packets. To ensure the validity of $PK_S$, $S$ generates the warrant's digest (Line~2) and verifies its congruence with $A$'s public key ($PK_A$) (Line~3). The correctness of the congruence in Line~3 shows the involvement of $A$'s public key in generating the delegated public verification key ($PK_S$). We note that Lines~1-3 of Protocol~\ref{sign} need to be executed once. Thus, the cost of executing these steps is negligible when amortized over multiple signing operations. To sign a Data packet (message $M$), $S$ executes Lines~4-8 of Protocol~\ref{sign}--the signing process follows Schnorr signature. $S$ selects a random integer $r \in \bZ^{*}_Q$ and calculates its corresponding value $k$ (Lines~4-5). It then uses $k$ in generating the message digest $a$ (Line~6). Using the private signing key ($PR_S$), the digest ($a$), and the random integer ($r$), $S$ signs the message (Line~7) and stores the signature as a five-tuple ($<M,W,t,a,b>$) for the consumer's verification process.
\begin{algorithm}[!t]
\caption{Delegated Message Signing by Proxy $S$}
\label{sign}
\scriptsize
\begin{algorithmic}[1]
\REQUIRE{$\bG_S$, $H()$, $M$, and $<PR_S, W, t>$.}
\ENSURE{Signed Message.}
\STATE Calculate proxy $S$ public key $PK_S = g^{PR_S} \in \bZ^{*}_P$.
\STATE Generate $W_h = H(W||t)$.
\IF {$\big(PK_S \equiv (PK_A^{W_h} \times t) \in \bZ^{*}_P \big)$}
\STATE Select $r = ZQrand()$.
\STATE Calculate $k = g^r \in \bZ^{*}_P$.
\STATE Generate $a = H(M||k) \in \bZ_Q$.
\STATE Calculate $b = \big(r - (PR_S \times a)\big) \in \bZ_Q$.
\STATE Store $<M,W,t,a,b>$.
\ELSE
\STATE Fail.
\ENDIF
\end{algorithmic}
\end{algorithm}
%
%

\begin{algorithm}[!b]
\caption{Signature Verification by Consumers}
\label{verify}
\scriptsize
\begin{algorithmic}[1]
\REQUIRE{$\bG_S$, $H()$, and $<M,W,t,a,b>$.}
\ENSURE{Verification Success / Fail.}
\STATE Generate $W_h = H(W||t)$.
\STATE Generate verification key $y \equiv (PK_A^{W_h} \times t) \in \bZ^{*}_P$.
\STATE Calculate $k_v = (g^b \times y^a) \in \bZ^{*}_P$.
\STATE Calculate $a_v = H(M||k_v) \in \bZ_Q$.
\IF {$(a == a_v)$}
\STATE Success.
\ELSE 
\STATE Fail.
\ENDIF
\end{algorithmic}
\end{algorithm}

\subsection{Signature Verification}

Protocols~\ref{generate} and~\ref{sign} enable consumers to validate the proxy's signatures and the delegation authorization, ensuring that $S$ is certified by $A$. Protocol~\ref{verify} details the verification process by accepting the Schnorr group ($\bG_S$), the hash function ($H()$), and the five-tuple generated by $S$ ($<M,W,t,a,b>$).

The consumer, verifying the \selProxy's signature, generates the warrant's digest ($W_h$) using warrant $W$ and $t$ from the signature (Line~1). The consumer then uses $W_h$ and $A$'s public key ($PK_A$) to derive the signature verification key ($y$). The amortized cost of extracting the signature verification key ($y$) will be negligible as Lines~1-2  will be executed once for a set of signature verification operations. 
After extracting $y$, the consumer executes Lines~3-4, which refer to the conventional Schorr signature verification process. Following Lines~5-9, the consumer accepts the signature if the received signature ($a$) matches the one that it generates ($a_v$) or rejects, otherwise.

%% file: Sections/design.tex
\section{Anonymous Data Uploading Design}
\label{sec:design}

In this section, we present the data uploading mechanism of \sol, so that data produced in a censoring network can become available to consumers outside of this network.

\subsection{Evading the Censoring Network}
\label{sec:p2p}



\subsubsection{Data sharing initialization} 

 The producer selects \collabPeers as a subset of the overall peers. The \collabPeers participate in uploading the producer's data outside of the censoring network. Subsequently, the producer creates uploading metadata 
for each \collabPeer, consisting of a symmetric key for the secure communication between the \collabPeer and the producer and the data pieces that the \collabPeer will forward to the proxies. 

\begin{figure}[!t]
\centering
\includegraphics[width=1\columnwidth]{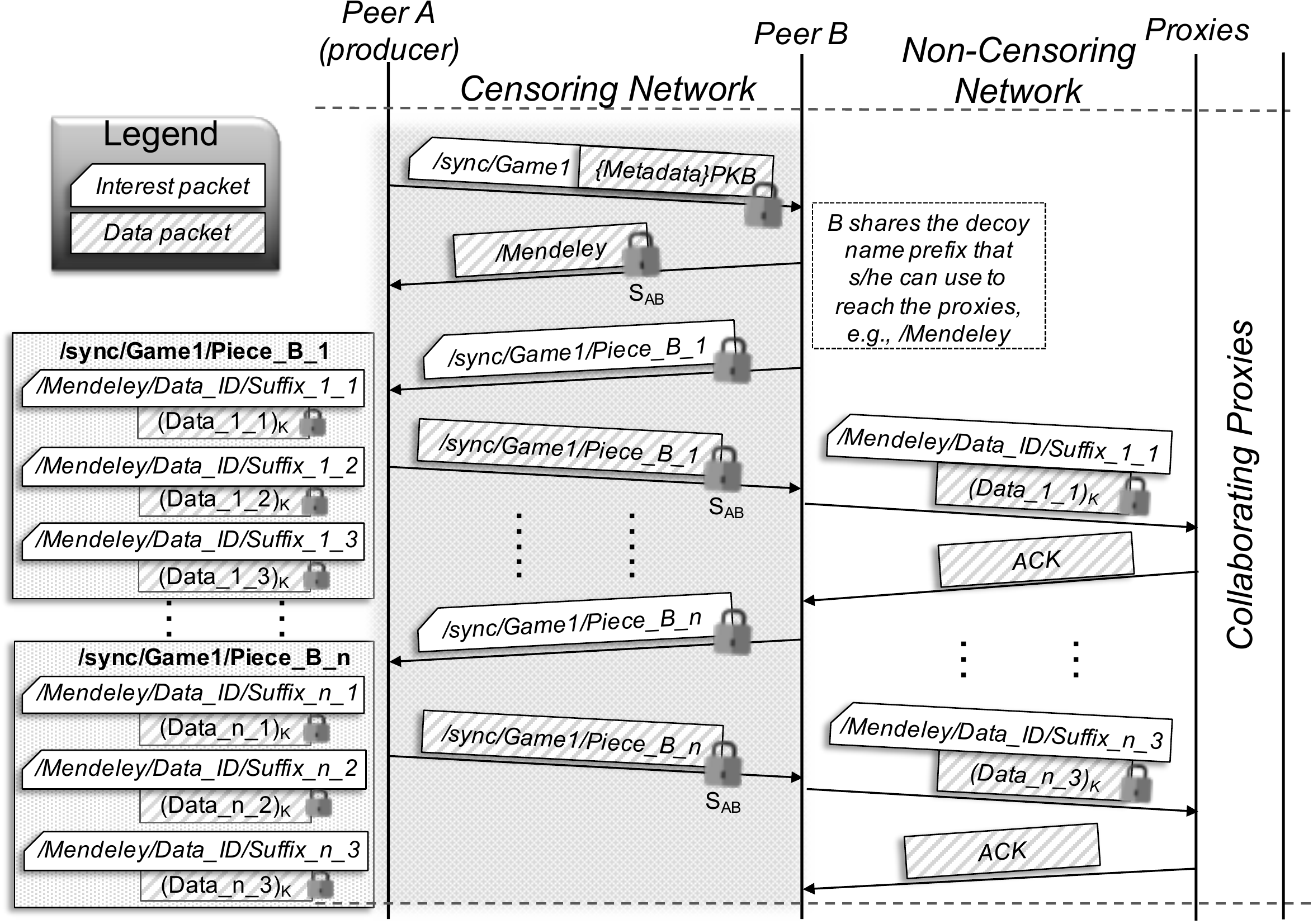}
\vspace{-0.7cm}
\caption{A data uploading example: peer A (producer) shares uploading metadata with peer B. Subsequently, peer B requests the data pieces specified in the metadata using the \pull communication mode. B forwards on behalf of the producer requests with decoy name prefixes contained in the received data pieces to \collabProxies. These requests carry the data to be made available outside of the censoring network in an encrypted format.}
\label{fig:sol}
\end{figure}

Fig.~\ref{fig:sol} illustrates a scenario, where the producer (peer A) distributes a subset of the total data to peer B, who will forward it to the proxies. The uploading metadata \\ $[Uploading\_MetaData]_{PK_B}$ sent from peer A to B contains the symmetric key $S_{AB}$ and a list of data pieces \name{/sync/Game1/Piece\_B\_1} $\cdots$ \name{/sync/Game1/Piece\_B\_k} that B should forward to the proxies. 
The metadata and the data pieces are named under a multicast synchronization (``sync'' for short) prefix used by a multi-party application (\eg gaming) allowed to operate in the censoring network. This prefix masquerades the producer's prefix, so that it stays anonymous. 


The producer sends uploading metadata to each of the \collabPeers (the metadata is encrypted using the receiving \collabPeer's public key) through the multicast synchronization channel. 
Thus, all peers in the multicast group will receive the metadata.
However, only the peer with the corresponding private key (peer B in Fig.~\ref{fig:sol}) will be able to decrypt the metadata and access the names of the data pieces that will be uploaded on behalf of the producer. In response, peer B will encrypt and send a decoy prefix (\eg \name{/Mendeley}) back to A. 
Each data piece listed in the uploading metadata will contain one or more Data packets named by A under B's decoy prefix. Each of these packets will contain the data to be anonymously uploaded to the proxies. Once B receives a data piece, it will decapsulate the contained packets and forward them towards the proxies outside of the censoring network as we further discuss in Section~\ref{subsubsec:available}.




\subsubsection{Data sharing modes among peers} 

Upon receiving the uploading metadata, a \collabPeer will request the data pieces specified in the metadata from the producer. We refer to this data sharing mode as \pull. These requests will be received by all peers in the multicast group, but only the producer has and will be able to provide the requested data.

Requests for data pieces may be intercepted by \censorNodes aiming to prevent the data from leaving the censoring network. The \censorNodes may receive requests sent by \collabPeers and reply with bogus pieces. In the example of Fig.~\ref{fig:sol}, B can detect a received bogus piece after trying to decrypt it using $S_{AB}$ (shared symmetric key between A and B).
As a result, B will request such pieces multiple times, alerting the producer that it has not received the legitimate pieces. Once the producer receives a certain number of consecutive requests for the same piece, the anti-censorship mode (\push data sharing mode) will be triggered. Under the \push mode, A will attach a piece requested multiple times onto an Interest and send (``push'') it through the multicast channel to B (encrypted with B's public key).





\sol features an adaptive communication mode (\hybrid data sharing mode) that operates under the \pull mode as long as no suspicious censorship activities are detected by the producer, while switching to the \push mode when \sol detects censoring activities. This adaptation will happen by the producer independently for each \collabPeer, 
since \censorNodes may be closer to the producer (thus being able to block requests) than only certain \collabPeers. 
To this end, the producer maintains a status for each \collabPeer and monitors the delivery progress of corresponding pieces.

\subsubsection{Making data available outside of the censoring network} 
\label{subsubsec:available}

As illustrated in Fig.~\ref{fig:sol}, once a \collabPeer receives and decrypts a data piece from the producer, this piece may contain one or more requests (Interests) for a decoy prefix. These requests carry (``hide'') the data to be uploaded in an encrypted format. As we mentioned in Section~\ref{subsec:assumptions}, given the pull-based nature of NDN communication, where data can be retrieved only after the reception of a request, access to the censoring network from the outside world may be easily restricted by the censor. 
To evade censorship and make the data available outside of the censoring network, the \collabPeers send the requests, found in the received pieces, towards the proxies. Due to their decoy name prefixes, these requests will be forwarded outside of the censoring network. 

\subsection{Data Gathering and Reconciliation}
\label{subsec:gathering}

As we explained in Section~\ref{sec:security}, the producer generates and shares with the \selProxy a {\it Data ID} random string. This is included in the names of the requests sent from the \collabPeers to the \collabProxies and is used to signal the \selProxy that these requests carry data belonging to a particular data collection. The \selProxy shares the {\it Data ID} value with all \collabProxies, instructing them to forward all the packets they receive and that contain this value in their names to the \selProxy. For instance, Fig.~\ref{fig:sol} illustrates that peers A and B agreed to use \name{/Mendeley} as the decoy prefix, while the requests sent to the \collabProxies have a name prefix \name{/Mendeley/Data\_ID}. Collaborating proxies receiving Interests for \name{/Mendeley} followed by \name{/Data\_ID} will forward them to the \selProxy. The suffix of the names can be selected by the producer based on the naming patterns of legitimate applications that use the decoy prefix, maximizing the resemblance between these requests and legitimate requests for the decoy prefix.

The requests received by the \collabProxies carry the data to be uploaded in an encrypted format, 
however, the \selProxy is the only entity that can decrypt this data, since it possesses the symmetric key K shared by the producer during the secure delegation process (Fig.~\ref{fig:phase1}). As a result, only the \selProxy can gather all the data, decrypt it, and reconcile the original data collection generated by the producer. The reconciled data will be published by the \selProxy to consumers under the name instructed by the producer during the secure delegation process (Fig.~\ref{fig:phase1}). 



%% file: Sections/eval.tex
\section{Evaluation}
\label{sec:eval}

In this section, we present our evaluation study under two setups. We first implement and evaluate our proxy signature design on different hardware platforms. We then implement \sol and perform network simulations, so that we can scale our study to large network topologies. Finally, we compare \sol to a design based on onion routing~\cite{goldschlag1999onion}. 


\subsection{Evaluation Setup}
\label{subsec:setup}

To evaluate the security delegation phase (Section~\ref{sec:security}), we implemented the proxy signature~\cite{AboYou12} and Schnorr signature~\cite{Sch91} mechanisms using the Charm-Crypto library~\cite{charm13}. We developed the proxy signature generation (Protocol~\ref{generate}), the proxy signing (Protocol~\ref{sign}), and the proxy verification (Protocol~\ref{verify}) protocols. We also implemented Schnorr message signing and signature verification as our comparison baseline. We benchmarked these protocols on three platforms: (i) a Raspberry Pi~4 with an ARMv7 processor and 4GB of RAM running Raspbian~10; (ii) a laptop with a 2.20GHz Intel Core-i7 processor and 4GB of RAM running an Ubuntu~16 Virtual Machine (VM); and (iii) a desktop class server with a 3.60GHz Intel Xeon processor and 16GB of RAM running Ubuntu~18. The results are averaged over 500 runs.

We use ndnSIM~\cite{ndnSIM}, the de-facto NDN network simulator, 
to implement and evaluate \sol 
based on a Rocketfuel topology (AS1221) with 278 routers and 731 links~\cite{spring2002measuring}. We connect collaborating peers and censoring nodes to this topology by creating links to randomly selected routers. We randomly attach five proxies to the topology, while ensuring that the distance between each proxy and the closest peer is at least five hops, so that each proxy is out of the censoring network. A file of size 100MB is generated by a producer (randomly selected among the peers) and is sent towards the proxies. 
Finally, we implemented a design based on onion routing~\cite{goldschlag1999onion} to compare with \sol. 
The realization of such an onion-based routing design in NDN is a challenge on its own, since NDN is fundamentally different than TCP/IP. To this end, in this paper, we randomly selected three onion routers and incorporated benchmarked encryption/decryption times of onion encryption operations for each onion router. For simplicity, we did not consider the time for the selection of onion routers and key exchanges. 
The results are averaged over ten runs.



\noindent \textbf {Evaluation metrics:}
We consider the following metrics: 

\begin{enumerate} [wide, labelwidth=!, labelindent=0pt, nosep]

\item {\em Run time of proxy and Schnorr signing and verification}: the time needed to perform the signing and verification operations on different hardware platforms. Proxy signing includes the time for proxy key derivation and Schnorr signature. Similarly, the proxy verification run time includes the time for proxy key derivation and Schnorr signature verification.

\item {\em Data distribution success rate}: the percentage of the total data that was successfully uploaded to the proxies.

\item {\em Data publication delay}: the time elapsed between the producer generating the data and the completion of the reception of all the data by the proxies.

\item {\em End-to-end per packet delay}: the time elapsed between starting the data uploading process for each Data packet and the reception of each packet by a proxy.

\item {\em Normalized overhead}: the ratio between the volume of overhead traffic (multicast communication, metadata exchanges, peer-to-peer data sharing) and the volume of the data to be uploaded from the producer to the proxies. We further normalize the overhead based on the traffic volume generated by the \pull mode. To this end, the \pull mode will, by definition, result in normalized overheads of value 1.

\end{enumerate}


\subsection{Evaluation Results}

\begin{figure}[t]
	\centering
	\vspace{-0.3cm}
	\includegraphics[width=0.85\linewidth]{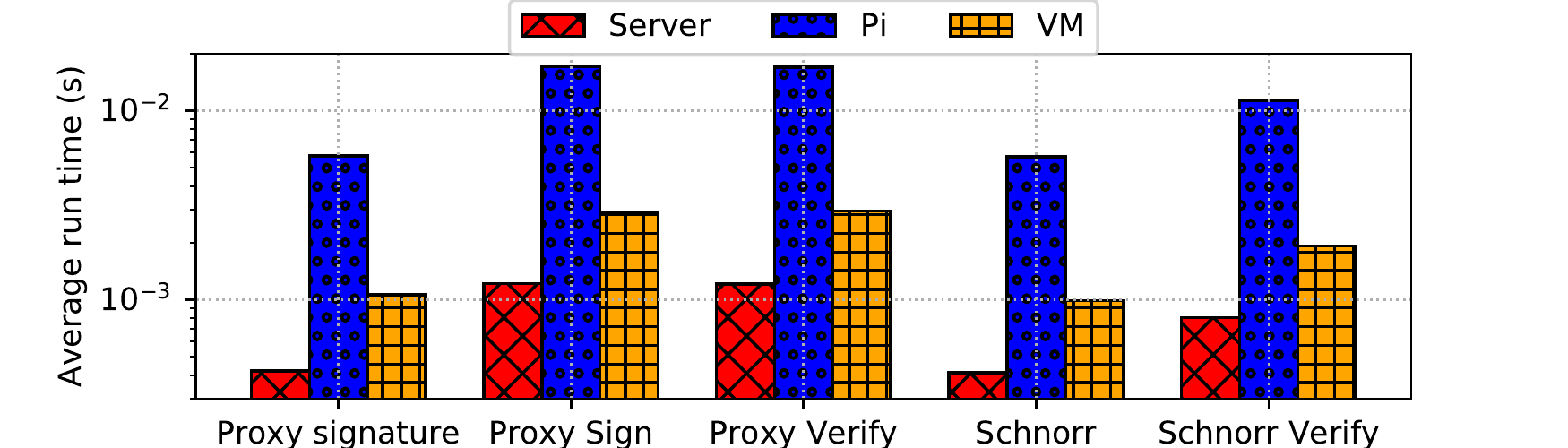}
	\vspace{-0.3cm}
	\caption{\label{Figures:security} Proxy and Schnorr signature implementation across different platforms. Run times are shown in log-scale.}
	\vspace{-0.1cm}
\end{figure}

\noindent \textbf{Run time of proxy and Schnorr signing and verification:} As shown in Fig.~\ref{Figures:security}, the proxy signature generation, executed by the producer, does not incur considerable delay even when running on a constrained device (6ms on a Raspberry Pi). The proxy signing process results in run times of about 3$\times$ higher than the run times for Schnorr signature on all platforms. The additional cost is attributed to the generation of the corresponding public key (for signature verification) and matching this key against the producer's public key (Lines~1-3 of Protocol~\ref{sign}). Similarly, the proxy verification results in run times of about 1.5$\times$ higher than Schnorr signature verification on all platforms due to the proxy's public key derivation (Lines~1-2 of Protocol~\ref{verify}). We emphasize that the key derivation and comparison (Protocols~\ref{sign} and~\ref{verify}) are executed only once per uploading session, incurring a negligible cost when amortized over multiple signing and signature verification operations.

\begin{figure}[t]
     \centering
     \vspace{-0.4cm}
     \includegraphics[width=0.6\columnwidth]{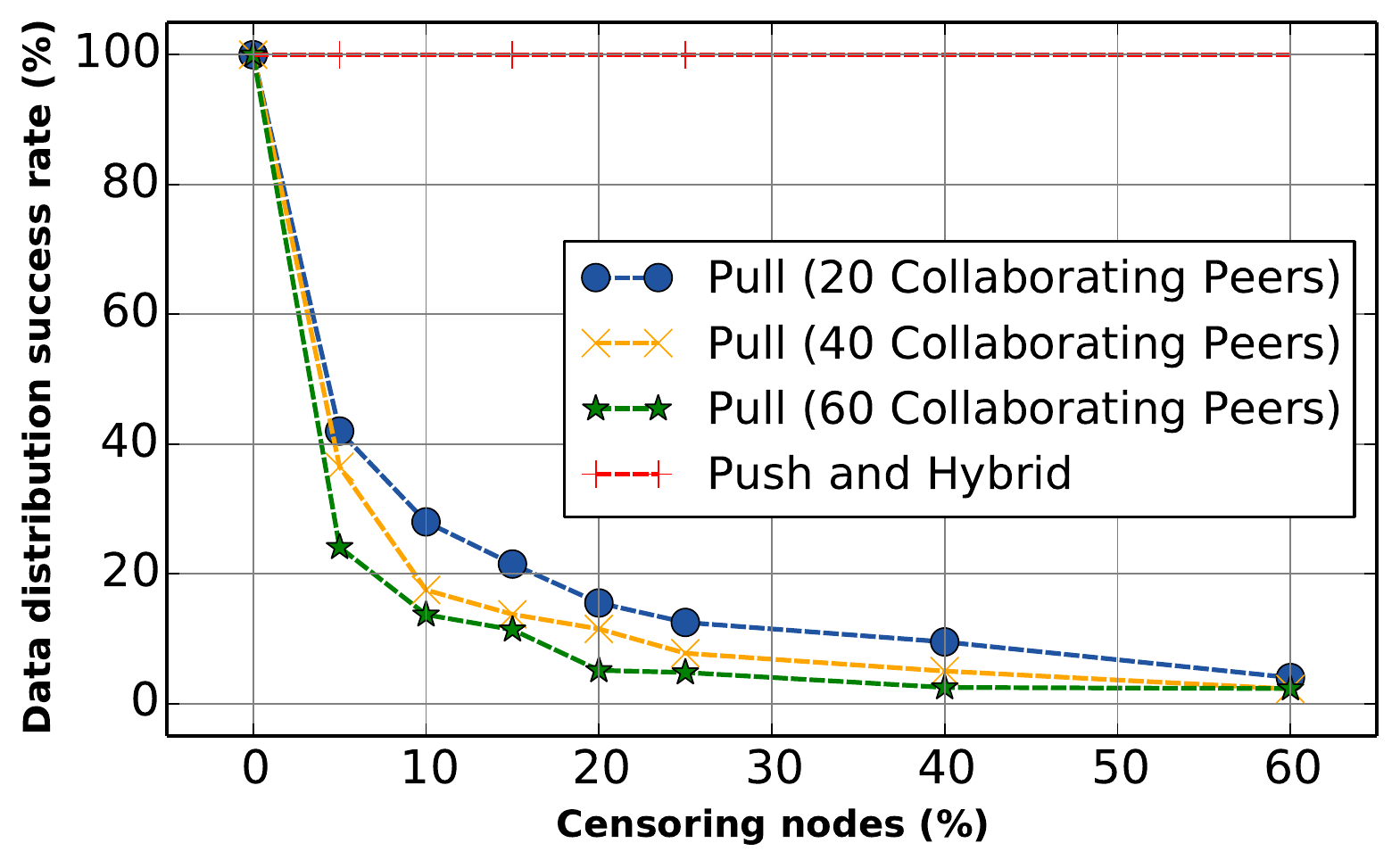}
     \vspace{-0.45cm}
     \caption{Success rate of different \sol data sharing modes (\pull, \push, and \hybrid). \push and \hybrid are aggregated into a single line as their success rates overlap.}
     \label{fig:success-rate}
     \vspace{-0.1cm}
\end{figure}

\begin{figure*}[t]
\vspace{-0.45cm}
    \captionsetup[subfigure]{aboveskip=-0.00000000001pt,belowskip=-0.00000000001pt}
	\centering
	\begin{subfigure}{0.33\textwidth}
		\centering
		\includegraphics[width=1.03\textwidth]{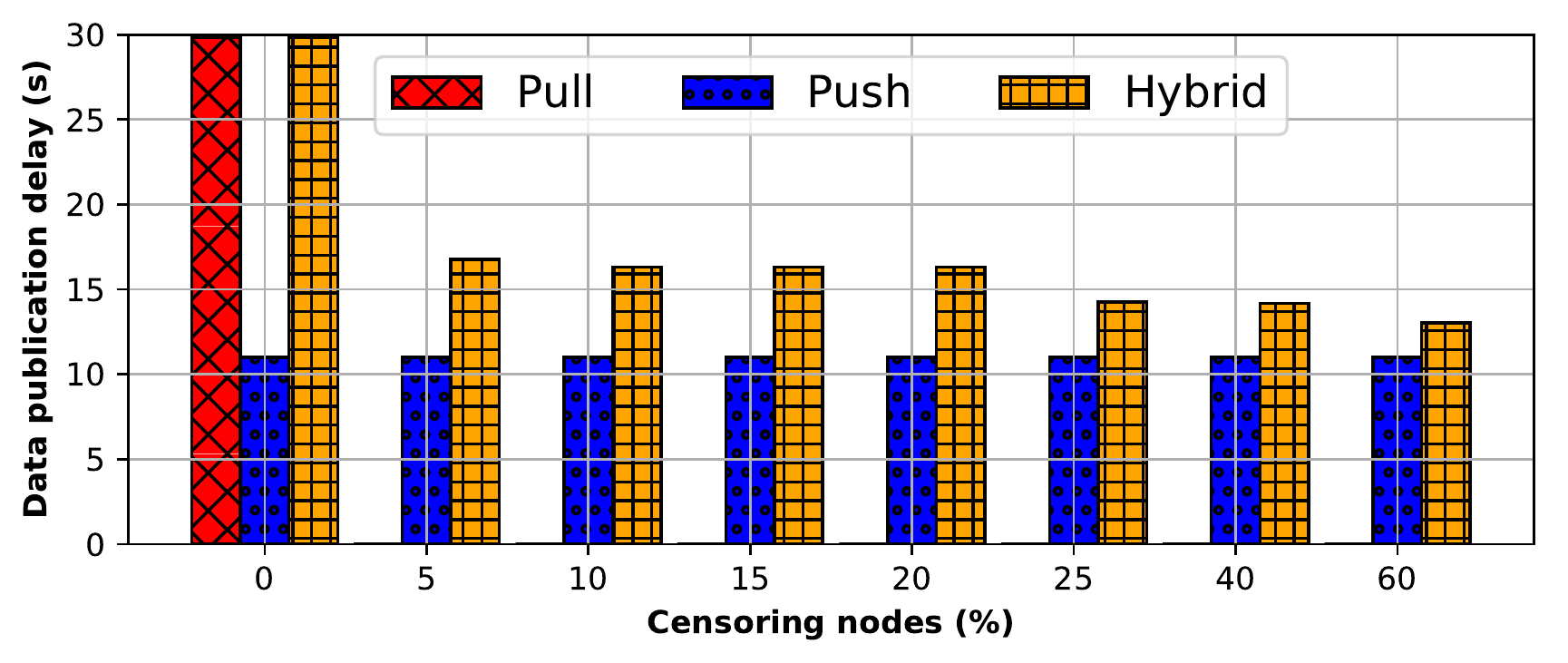}
		\caption{20 \collabPeers}
		\label{Figure:20peersdelay}
	\end{subfigure}
	\begin{subfigure}{0.33\textwidth}
		\centering
		\includegraphics[width=1.03\textwidth]{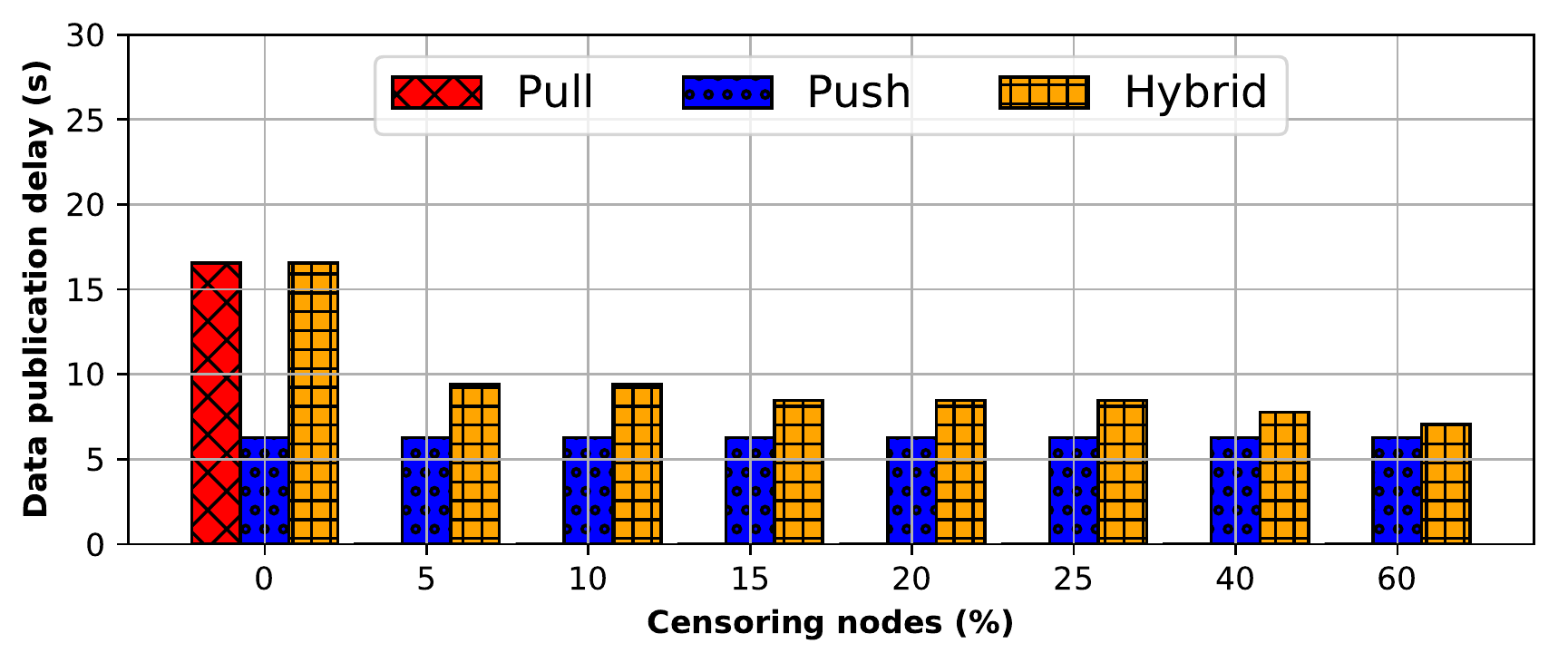}
		\caption{40 \collabPeers}
		\label{Figure:40peersdelay}
	\end{subfigure}
	\begin{subfigure}{0.33\textwidth}
		\centering
		\includegraphics[width=1.03\textwidth]{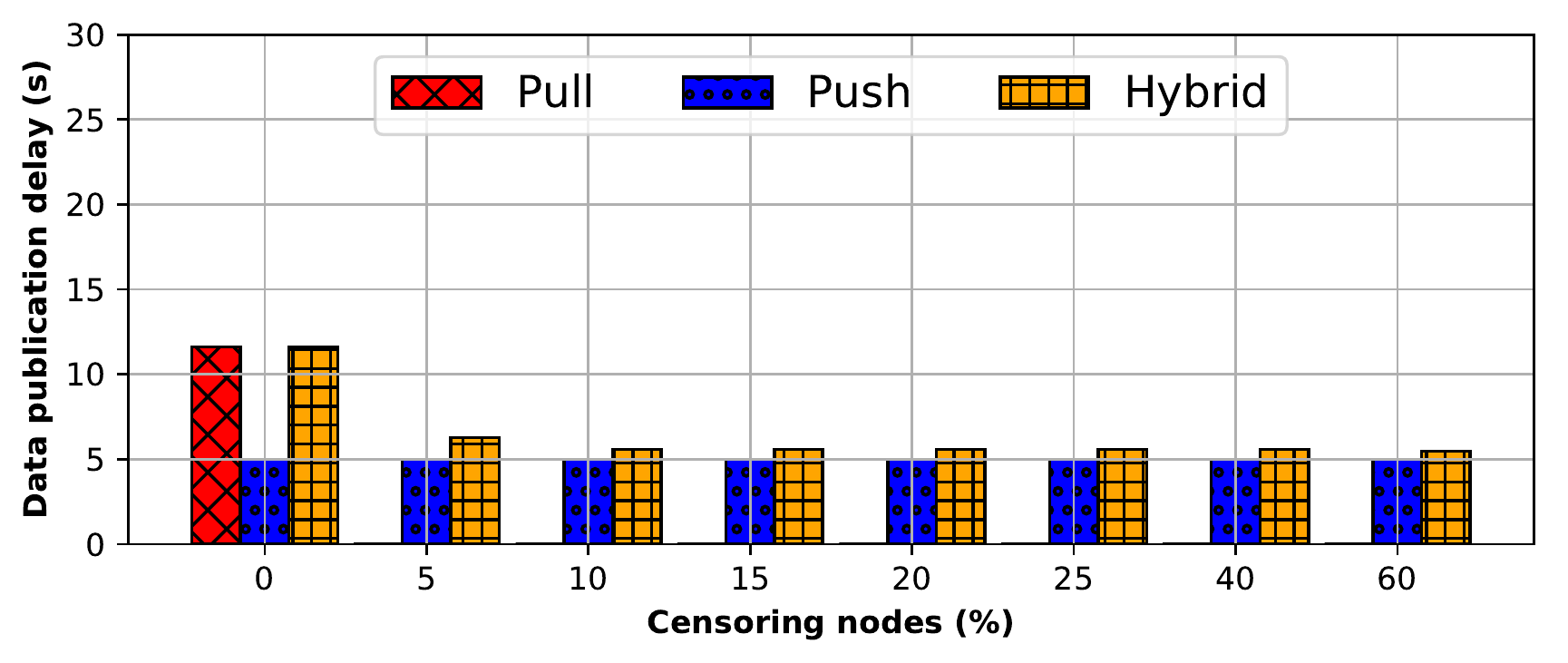}
		\caption{60 \collabPeers}
		\label{Figure:60peersdelay}
	\end{subfigure}
	\vspace{-0.4cm}
	\caption{\label{Figures:delay} Data publication delay for varying numbers of \collabPeers and percentages of \censorNodes. Results for \pull are omitted when it fails to successfully make all the data available outside of the censoring network.}
	\vspace{-0.2cm}
\end{figure*}

\begin{figure}[!th]
\vspace{-0.25cm}
\captionsetup[subfigure]{aboveskip=-0.00000000000000001pt,belowskip=-0.00000000000000001pt}
	\centering
	\begin{subfigure}[b]{0.49\columnwidth}
		\centering
		\includegraphics[width=1.03\textwidth]{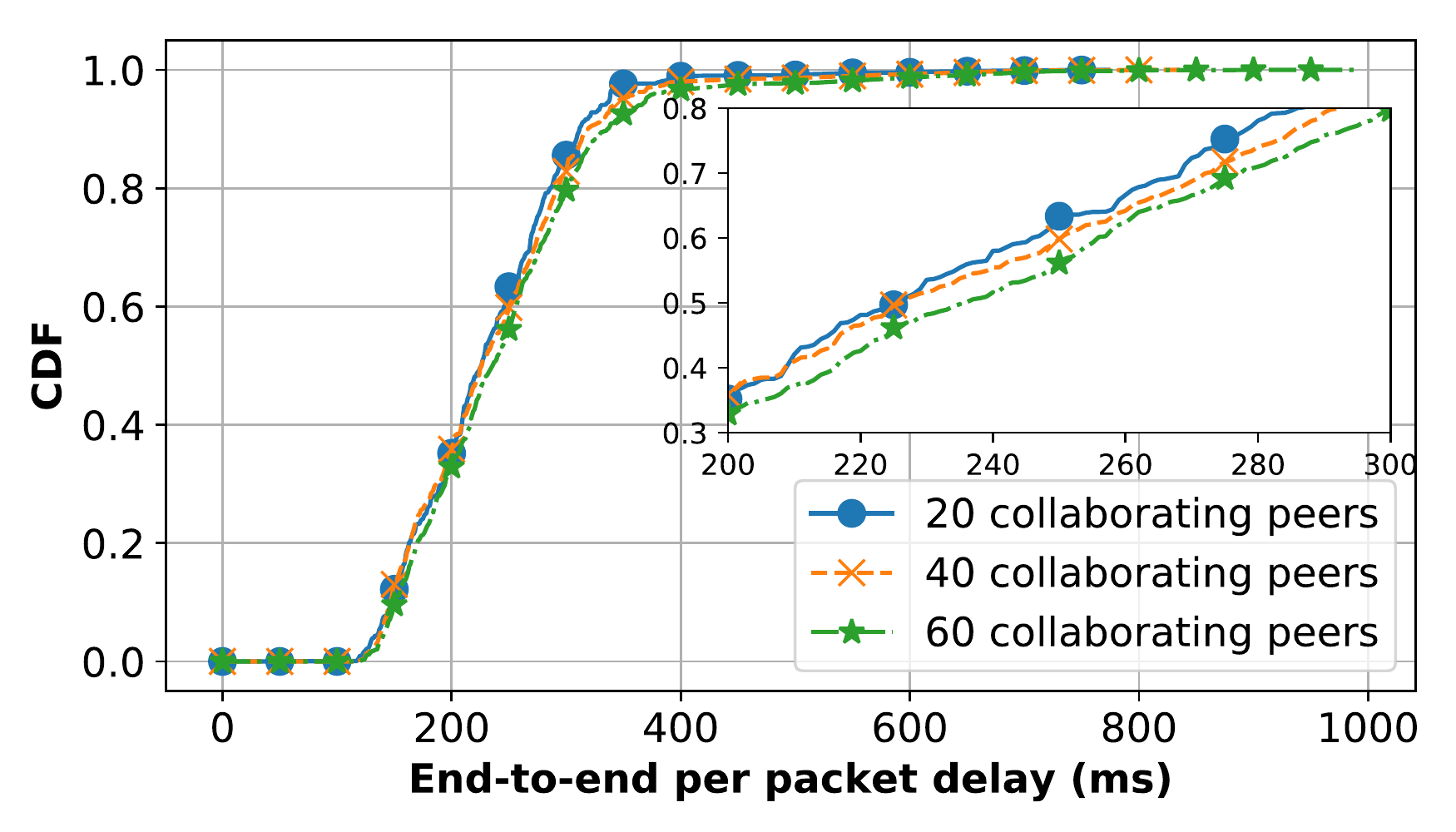}
		\caption{20\% censoring nodes (with 20, 40, and 60 \collabPeers)}
		\label{Figure:cdfa}
	\end{subfigure} \hfil
	\begin{subfigure}[b]{0.49\columnwidth}
	    \centering
	    \includegraphics[width=1.03\textwidth]{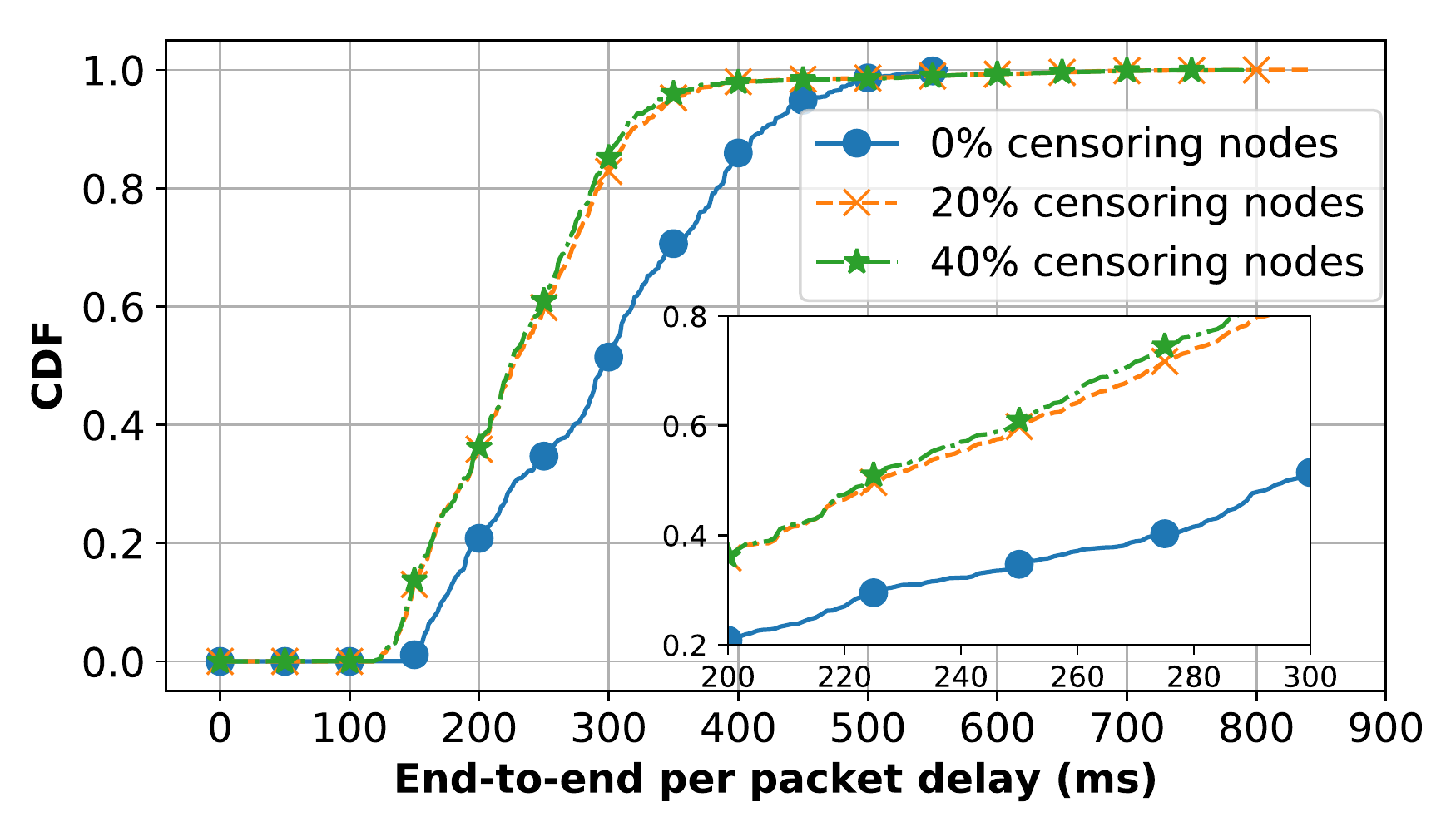}
	    \caption{40 \collabPeers (with 20\%, 40\%, and 60\% censoring nodes)}
	    \label{Figure:cdfb}
	\end{subfigure}\hfil
	\vspace{-0.3cm}
	\caption{\label{Figure:cdf}CDF of the end-to-end per packet delay. Markers do not represent actual data points, but are only used for better readability.}
	\vspace{-0.1cm}
\end{figure}

\begin{figure*}[t]
\vspace{-0.45cm}
    \captionsetup[subfigure]{aboveskip=-0.1pt,belowskip=-0.1pt}
	\centering
	\begin{subfigure}{0.325\textwidth}
		\centering
		\includegraphics[width=1.03\textwidth]{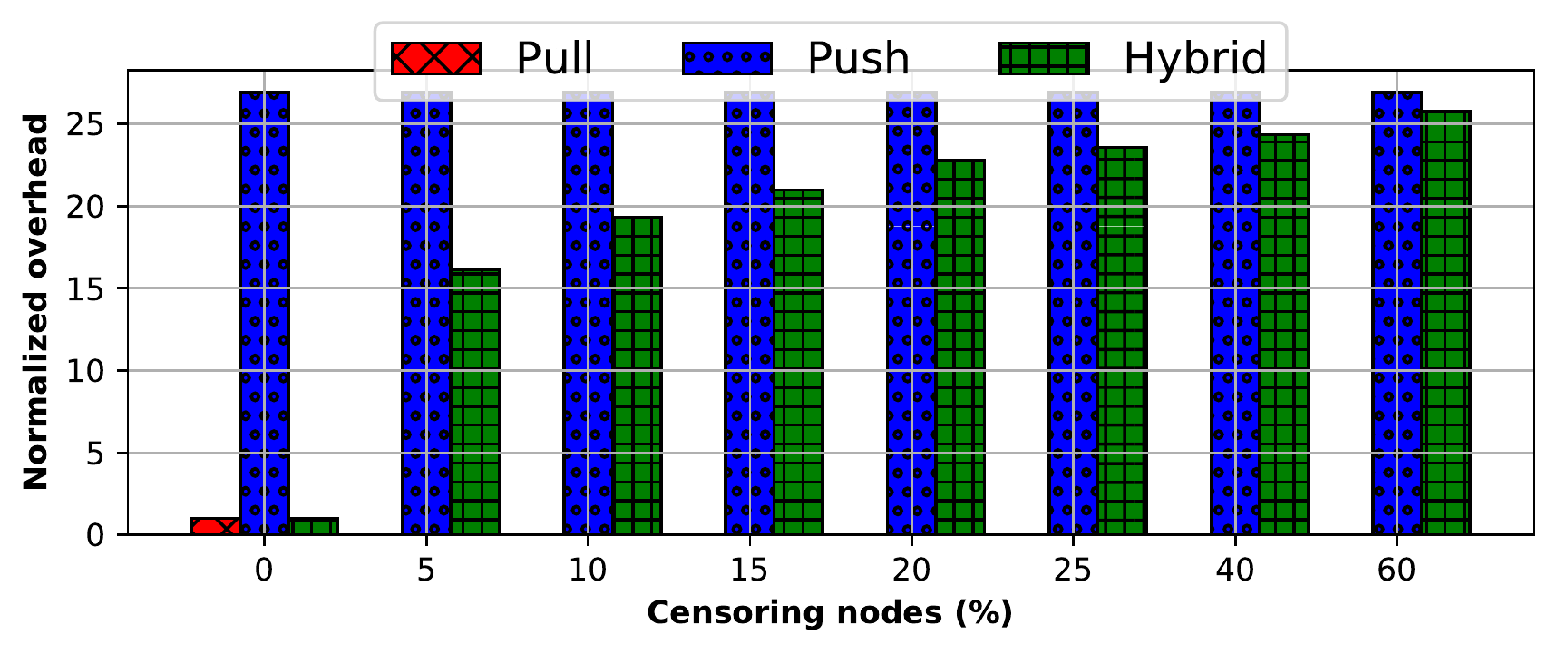}
		\caption{20 \collabPeers}
		\label{Figure:20peersoverhead}
	\end{subfigure}
	\begin{subfigure}{0.325\textwidth}
		\centering
		\includegraphics[width=1.03\textwidth]{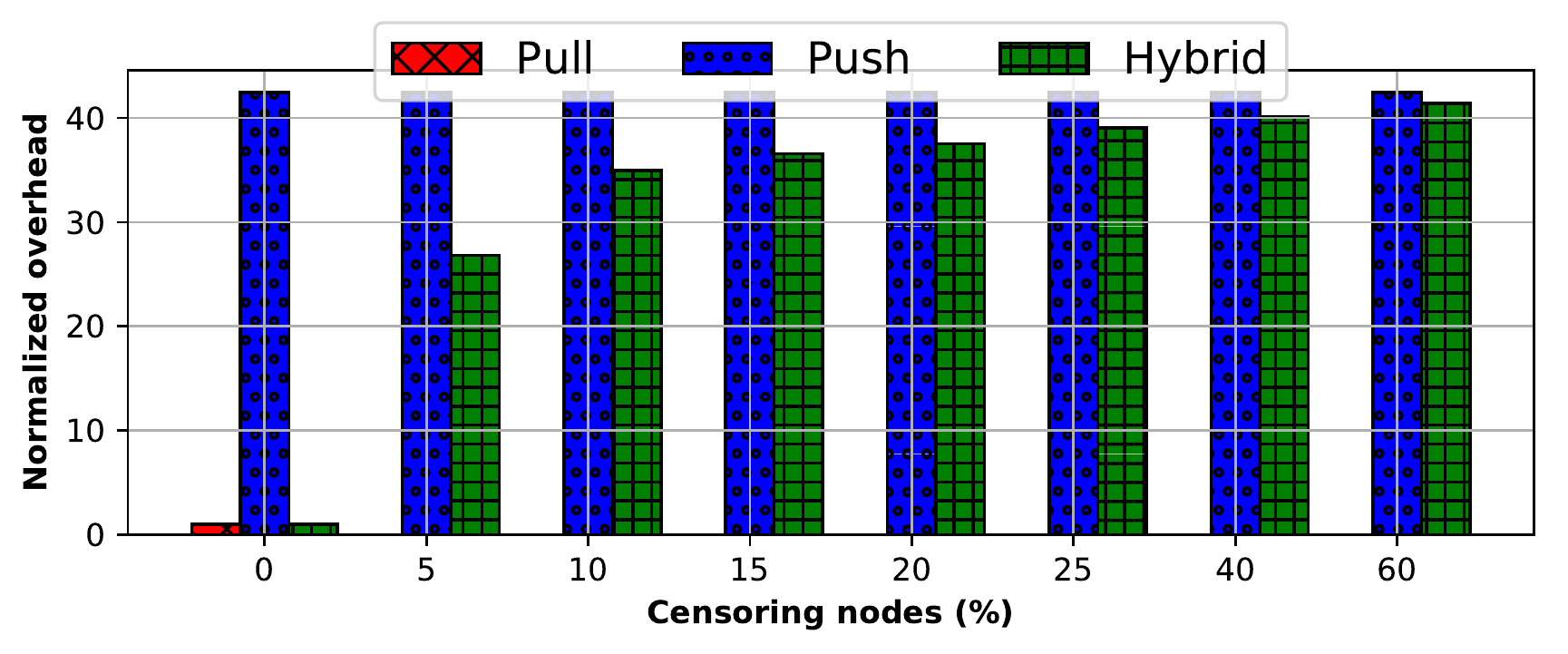}
		\caption{40 \collabPeers}
		\label{Figure:40peersoverhead}
	\end{subfigure}
	\begin{subfigure}{0.325\textwidth}
		\centering
		\includegraphics[width=1.03\textwidth]{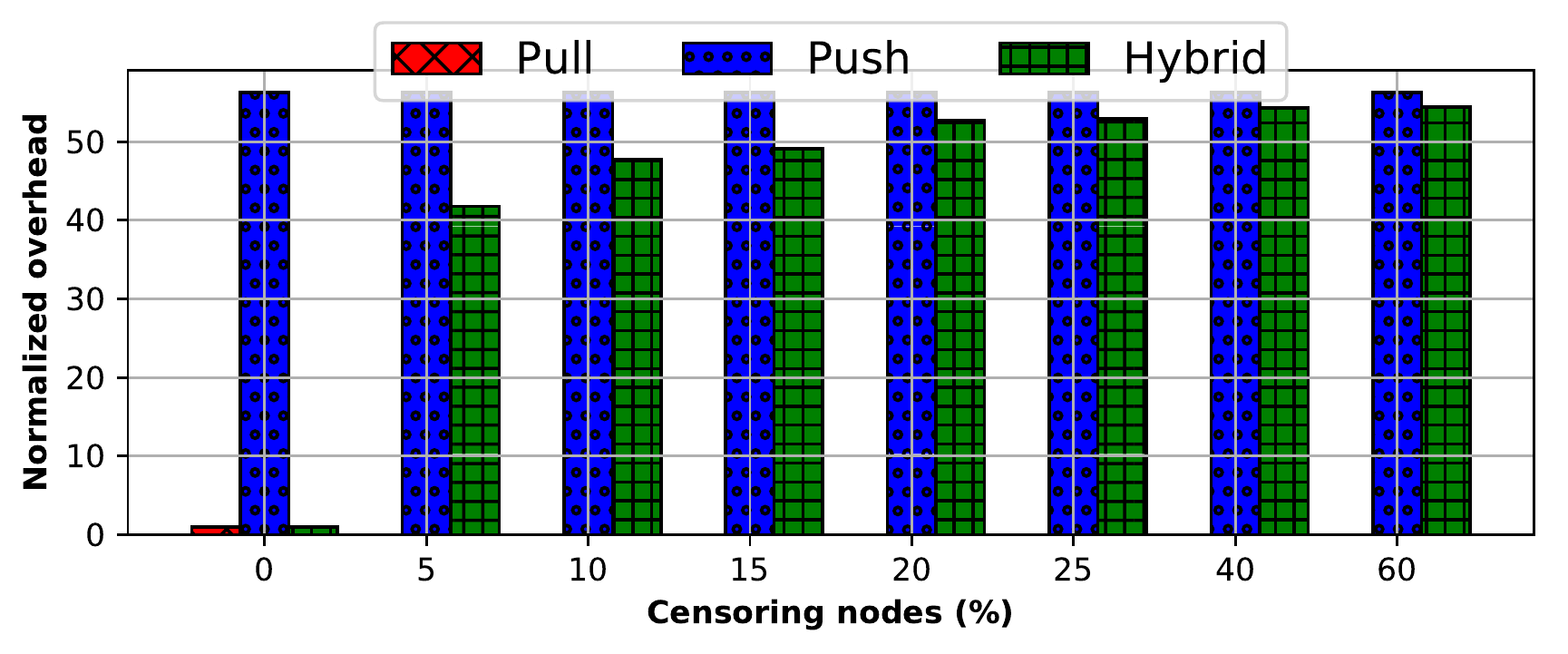}
		\caption{60 \collabPeers}
		\label{Figure:60peersoverhead}
	\end{subfigure}
	\vspace{-0.35cm}
	\caption{\label{Figures:overhead}Normalized overhead for varying numbers of \collabPeers and percentages of \censorNodes. Results for \pull are omitted when it fails to successfully make all the data available outside of the censoring network.}
    \vspace{-0.45cm}
\end{figure*}

\noindent \textbf{Data distribution success rate:} In Fig.~\ref{fig:success-rate}, we present the data distribution success rate. Our results show that \hybrid and \push modes successfully upload all the produced data to the proxies. On the other hand, in the case of \pull, censoring nodes are able to intercept the requests for data pieces sent by the \collabPeers towards the producer. To this end, \collabPeers will not be able to receive and distribute the data towards the proxies. The actual success rate values depend on the actual placement of the censoring nodes. However, the random placement in our experiments shows that even for small percentages of \censorNodes (5\% to 10\%), the success rate of \pull degrades considerably, since the majority of \collabPeers is blocked by \censorNodes as presented in Table~\ref{table:blockedpeers}. 
Specifically, 58-75\% and 72-86\% of the \collabPeers are blocked for 5\% and 10\% of \censorNodes respectively. As the percentage of \censorNodes increases, up-to 96-97.75\% of the \collabPeers may be blocked. Nevertheless, even in such cases, \sol successfully uploads all the produced data to the proxies.


\begin{table}[t]
\vspace{-0.3cm}
\caption{Percent of \collabPeers blocked by \censorNodes.} 
\vspace{-0.35cm}
\centering
\resizebox{\columnwidth}{!}{
\begin{tabular}{|c|c|c|c|c|c|c|c|c|}
\hline
\multirow{2}{*}{
\textbf{Collaborating Peers}} & \multicolumn{8}{c|}{\textbf{Censoring Nodes (\%)}}                                                                          \\ \cline{2-9}
\textbf{}                    & \textbf{0\%} & \textbf{5\%} & \textbf{10\%} & \textbf{15\%} & \textbf{20\%} & \textbf{25\%} & \textbf{40\%} & \textbf{60\%} \\ \hline
\textbf{20}                  & 0\%          & 58\%         & 72\%          & 78.5\%        & 84.5\%        & 87.5\%        & 90.5\%        & 96\%          \\ \hline
\textbf{40}                  & 0\%          & 63.5\%       & 82.5\%        & 86.25\%       & 88.5\%        & 92\%          & 95\%          & 97.75\%       \\ \hline
\textbf{60}                  & 0\%          & 75\%         & 86\%          & 88.5\%        & 94.8\%        & 95\%          & 97.5\%        & 97.6\%        \\ \hline
\end{tabular}
}
\label{table:blockedpeers}
\vspace{-0.1cm}
\end{table}

\noindent \textbf{Data publication delay:} In Fig.~\ref{Figures:delay}, we present the results of the average data publication delay for \pull, \push, and \hybrid. 
Our results indicate that the data publication delay for \push is the lowest and it does not increase as the number of censoring nodes increases, since the data pieces are pushed to all nodes including the censors, while only the \collabPeers can decrypt these pieces. 
\pull's performance suffers in the presence of censoring nodes, even if their number is relatively small (\eg 5\% or 10\% of the number of \collabPeers). When the percentage of censoring nodes increases from 0\% to 5\%, \pull fails to distribute all data pieces among the \collabPeers. \hybrid, however, successfully adjusts to the censoring nodes that intercept the data pieces, switching to the \push mode. Our results show that as the percentage of censoring nodes increases, \hybrid switches from the \pull to the \push mode sooner during the data publication process, thus \hybrid's data publication delay converges towards the delay of \push. For all the modes, the data publication delay decreases as we increase the number of \collabPeers due to the fact that more peers upload the data in parallel. 

Further analysis of our results indicated that 3-8\% of the data publication delays are spent on sharing the metadata between producers and peers, 47-56\% on sharing the data pieces between producers and peers, and 36-50\% on sending the actual data from the \collabPeers to the proxies. Note also that the \hybrid mode results in 1.5-2.1$\times$ higher publication delays than uploading the data from the producer to the closest proxy directly over the shortest network path.


\noindent \textbf{End-to-end per packet delay:} Fig.~\ref{Figure:cdf} presents the CDF of the per packet delay. Fig.~\ref{Figure:cdfa} shows that for varying numbers of \collabPeers (same percentage of censoring nodes), 40\% and 80\% of the data is uploaded in less than 200ms and 300ms respectively. The per packet delay slightly increases with the number of \collabPeers, since these peers may be further away from the producer, thus the data pieces travel longer distances to reach them. Fig.~\ref{Figure:cdfb} shows that the per packet delay decreases as the percentage of censoring nodes increases, since more pieces are blocked, thus \sol switches from \pull to \push sooner during data publication.

\noindent \textbf{Normalized overhead:} Fig.~\ref{Figures:overhead} shows the normalized overhead results. The overhead for \pull is equal to 1, since it acts as the normalization factor. \push results in the highest overheads, since the data pieces are attached onto Interests pushed towards \collabPeers. \hybrid successfully copes with the interception of data pieces by censoring nodes, achieving overheads in the range between \pull and \push. It converges to the overhead of \pull when no or a few censoring nodes exist and to the overhead of \push as the number of censoring nodes increases. As the number of \collabPeers increases, the overhead for \push and \hybrid increases, since the size of the peer multicast group increases. 

\noindent \textbf{Comparison to an onion routing based design:} Compared to a design based on onion routing, \sol achieves 1.33-4.05$\times$ lower data publication delays, since it does not require multiple time-consuming layers of encryption/decryption. Depending on the placement of onion routers, \sol incurs roughly the same to up to 1.51$\times$ lower overheads for \pull and \hybrid compared to the onion routing based design when no censoring nodes exist. As we increase the number of censoring nodes and collaborating peers, the \hybrid mode of \sol incurs 1.21-2.05$\times$ higher overheads compared to the design based on onion routing, since: (i) it switches from \pull to \push sooner during data publication as we increase the number of censoring nodes; and (ii) the size of the peer multicast group increases as we increase the number of collaborating peers.








%% file: Sections/discussion.tex
\section{Security Analysis and Discussion}
\label{sec:discussion}
%
In this section, we discuss further security considerations and directions to extend the design of \sol.

%

\noindent \textbf{Censoring network nodes:} A censoring authority may deploy \censorNodes in the network, including routers and Deep Packet Inspection (DPI) proxy firewalls, to interrupt data publication or breach the producer's anonymity. A censoring router, due to its limited capability in processing Interest and Data packets beyond name matching, can randomly drop a subset of packets. Such an action will negatively impact peers that legitimately use allowed communication channels. 
Prior work has argued that censoring authorities avoid actions that result in high collateral damage~\cite{zolfaghari2016practical}. 

Malicious routers may redirect traffic portions to proxy firewalls for DPI. Such redirection in NDN is complicated due to the communication model symmetry. More importantly,
Interest and Data packets, although semantically rich, do not carry fine-grained information that is available in TCP/IP packets (\eg IP addresses and port numbers). We argue that in inspecting NDN packets, a proxy firewall can only use packet sizes, names, and signature related information. \sol limits the impact of these threats by: (i) uploading small-sized Data packets by attaching them to Interests to avoid traffic anomalies; (ii) revealing innocuous names (\ie used by traffic allowed in the censoring network); and (iii) referring to anonymous certificates in the signature related information~\cite{ndn2015ndn} of data pieces to prevent the producer's linkability to the data. It will be computationally expensive for a censoring authority to verify the signatures of all data pieces. However, if DPI drops all pieces associated with anonymous certificates, \sol will switch from the \pull to the \push mode, piggybacking pieces onto Interests (typically not signed in most NDN applications).
 
%
%
\noindent \textbf{Censoring peers:} 
The censoring authority may deploy \censorNodes among peers. A censoring peer may intercept requests for data pieces from \collabPeers and reply with bogus pieces, which will consume PIT entries on routers and prevent the legitimate pieces from reaching the \collabPeers. The \hybrid mode of \sol thwarts this threat by switching to \push when such an event is identified. As we discussed in Section~\ref{sec:eval}, the \hybrid mode achieves 100\% data distribution success rates in the presence of censoring peers--even when 60\% of the peers are malicious. 

If \censorNodes are among the \collabPeers, these censoring \collabPeers can interrupt the communication by obtaining and dropping the producer's data pieces (blackhole attacks). Although we assumed that the \collabPeers are not malicious (Section~\ref{subsec:assumptions}), here we discuss directions to thwart such an attack. The first direction involves data replication. In a naive approach, the producer blindly replicates the data by communicating overlapping data portions to different \collabPeers. This increases the chances that the data will be received by legitimate \collabPeers, who will upload it towards the proxies. To minimize redundant data delivery, the producer can obtain the list of missing Data packets from the \selProxy and publish them through the \collabPeers that delivered previous packets. The producer identifies legitimate \collabPeers by tracking their success rates in delivering data to the \selProxy. Network coding techniques, such as Random Linear Network Coding~\cite{HoKoeMed03}, 
can also be employed to deliver linearly independent combinations of Data packets to the \selProxy, enabling efficient data reconciliation.

The second direction involves group-oriented cryptographic techniques such as attribute-based~\cite{goyal2006attribute} and broadcast~\cite{fiat1993broadcast} encryption. These techniques enable a group of \collabPeers to use their private keys to independently decrypt the same data piece delivered to them during the \push mode over the multicast communication channel. If, at least, one of the \collabPeers that can decrypt each data piece is legitimate, the data will be successfully uploaded to a proxy. 



%
%
\noindent \textbf{Producer and collaborating peer anonymity:} 
In our design, the producer includes its public key in the warrant, enabling consumers to verify the validity of the delegation in addition to the proxy's signatures. The producer's public key in the warrant may allow the censoring authority to identify the producer, compromising its anonymity. To cope with this threat, approaches that provide signature anonymity can be used, including attribute-based~\cite{ramani2019ndn}, ring~\cite{RivShaTau01}, and group signatures~\cite{Cha91}. The producer's anonymity can be also augmented through a transient key cryptosystem~\cite{Bra83}, an asymmetric key cryptosystem, in which the key pair is bound to a short time period rather than the owner's identity. Thus, a singed Data packet will be associated with a time (delegation initiation in \sol) rather than an identity. However, utilizing such a cryptosystem requires further considerations since private keys will be deleted after their short expiry time.

A malicious producer (deployed by the censoring authority)
may be able to infer the participation of \collabPeers in data uploading, compromising their privacy. Similar to the producer's anonymity, cryptosystems including ring, group, and attribute-based signatures can preserve peers' anonymity. Distributed anonymous reputation management mechanisms can also help peers make informed decisions about their participation in data uploading~\cite{WanCheMoh13}.



%
%
\noindent\textbf{Traffic analysis attacks:} The censoring authority may orchestrate traffic analysis attacks to infer communication patterns from encrypted traffic, aiming to breach the producer's anonymity. 
Note that data producers in \sol use legitimate communication channels 
to transfer their data to the collaborating peers and subsequently to the proxies. Leveraging such legitimate communication channels for distributing the data between the collaborating peers hides the producer's data and prevents the censoring authority from identifying a data upload attempt.
As described in Section~\ref{sec:p2p}, dispersing the producer's data across multiple collaborating peers allows each peer to obtain a small portion of the data--with potentially different sizes--from the producer. Peers can send requests for data to the producer such that the generated traffic follows the legitimate application distribution, making these requests indistinguishable from the traffic generated by the legitimate application. Each peer also uploads a small portion of the generated data to the proxies, thus preventing peers from sending abnormal amounts of data outside of the censoring network and creating traffic anomalies. 

To transfer the producer's data to proxies outside of the censoring network, the collaborating peers send requests containing portions of the data generated by the producer hidden in them. The censoring authority may attempt to orchestrate traffic correlation attacks by passively observing the packet sizes between the producer and the collaborating peers or between the collaborating peers and the proxies. However, the data producer can assign different portions of the data to collaborating peers, ensuring that traffic patterns between the producer and these peers are not identical. NDN also features variable size request packets, since such packets can carry an unbounded number of parameters (data of arbitrary sizes) as defined by the NDN packet format~\cite{ndn-parameters}. As a result, data producers can generate Interests of variable sizes that follow the packet sizes of legitimate applications.

%% file: Sections/conclusion.tex
\section{Conclusion and Future Work}
\label{sec:conclusion}

In this paper, we presented \sol, a framework for the anonymous publication of data from a censoring network to users outside of this network. \sol takes advantage of communication channels and applications that are allowed in the censoring network, maximizing the collateral damage for censoring authorities. By employing different data sharing modes, \sol can defend against censoring actions. Through a secure delegation mechanism, \sol enables proxies outside of a censoring network to make data available to users without compromising the producer's anonymity. In the future, we plan to: (i) implement a \sol prototype and evaluate it against other censorship circumvention solutions; and (ii) design mechanisms to defend against malicious collaborating peers and proxies.